\documentclass[aps,11pt,onecolumn,preprintnumbers,superscriptaddress,amsmath,amssymb,tightenlines,longbibliography,nofootinbib,notitlepage]{revtex4-1}
\usepackage{graphicx}
\usepackage{dcolumn}
\usepackage{bm}
\usepackage{multirow} 
\usepackage{comment}
\usepackage{url}
\usepackage{epsfig}
\usepackage{amsfonts}
\usepackage{amsmath}
\usepackage{bm}
\usepackage{graphicx}
\usepackage{color}
\usepackage{booktabs}
\usepackage{amssymb,amsmath,hyperref,slashed,cancel}
\usepackage[normalem]{ulem}
\usepackage[T1]{fontenc}
\usepackage{relsize}
\usepackage{gensymb}
\usepackage{fontawesome}
\usepackage{csvsimple}
\usepackage{pgfplotstable}
\usepackage{longtable}
\usepackage{xspace} 

\usepackage[capitalize]{cleveref}
\usepackage{siunitx}
\usepackage{xcolor}
\usepackage{titlesec} 
\usepackage{subfigure}
\usepackage[subfigure,titles]{tocloft} 
\allowdisplaybreaks

\definecolor{col1c100s}{rgb}{0.13,0.53,0.20}
\definecolor{col100c1s}{rgb}{0.40,0.80,0.93}

\definecolor{niceblue}{rgb}{0.117,0.5625,1.0}
\hypersetup{colorlinks,
			citecolor      = niceblue, 
			linkcolor       = niceblue, 
			urlcolor        = niceblue, 
			anchorcolor = niceblue}

\makeatletter
\renewcommand{\thesection}{\arabic{section}}

\renewcommand{\p@subsection}{}
\makeatother

\titleformat*{\section}{\centering\bfseries\scshape}
\titlelabel{\thetitle\quad}
\titleformat*{\paragraph}{\bfseries}
\titlespacing*{\paragraph}{0pt}{3.25ex plus 1ex minus .2ex}{1em}

\makeatletter
\def\l@subsubsection#1#2{}
\makeatother

\newcommand{\iso}[2]{{\ensuremath{{}^{#2}}\ensuremath{\rm #1}}}

\definecolor{nicegreen}{rgb}{0., 0.75, 0.46}

\newcommand{\nuance}{\texttt{NUANCE}\xspace}

\begin{document}

\preprint{CERN-TH-2022-151}

\title{{\Large More Ingredients for an Altarelli Cocktail at MiniBooNE}}

\author{Kevin J. Kelly}
\affiliation{Theoretical Physics Department, CERN, Esplande des Particules, 1211 Geneva 23, Switzerland}
\affiliation{Department of Physics and Astronomy, Mitchell Institute for Fundamental Physics and Astronomy, Texas A\&M University, College Station, TX 77843, USA}

\author{Joachim Kopp}
\affiliation{Theoretical Physics Department, CERN, Esplande des Particules, 1211 Geneva 23, Switzerland}
\affiliation{PRISMA+ Cluster of Excellence \& Mainz Institute for Theoretical Physics,
             Staudingerweg 7, 55128 Mainz, Germany}

\date{\today}

\begin{abstract}
The MiniBooNE excess persists as a significant puzzle in particle physics. Given that the MiniBooNE detector cannot discriminate between electron-like signals and backgrounds due to photons, the goal of this work is to study photon backgrounds in MiniBooNE in depth.  We first consider a novel single-photon background arising from multi-nucleon scattering with coherently enhanced initial or final state radiation. This class of processes, which we dub ``2p2h$\gamma$'' (two-particle--two-hole + photon) can explain ${\sim}40$ of the ${\sim}560$ excess events observed by MiniBooNE in neutrino mode. Second, we consider the background from neutral-current single-$\pi^0$ production, where two photons from $\pi^0\to\gamma\gamma$ decay are mis-identified as an electron-like shower. We construct a phenomenological likelihood that reproduces MiniBooNE's $\pi^0\to\gamma\gamma$ background faithfully. Even with data-driven background estimation techniques, we find there is a residual dependence on the Monte Carlo generator used. Our results motivate a reduction in the significance of the MiniBooNE excess by $0.4\sigma$.
\end{abstract}

\maketitle

\newcommand{\contentsname}{}
\setlength{\cftbeforesecskip}{3pt}
\setlength{\cftsecnumwidth}{2em}
\setlength{\cftsubsecindent}{\cftsecnumwidth}
\setlength{\cftsubsecnumwidth}{3em}
{
    \cftsetpnumwidth{4em}
    \tableofcontents
}

\clearpage

\section{Introduction}


At $4.8\sigma$, the MiniBooNE anomaly is currently the most statistically significant unexplained anomaly in neutrino physics. MiniBooNE -- a short-baseline neutrino oscillation experiment at Fermilab -- have observed an intriguing excess of events in their search for electron neutrino ($\nu_e$) appearance in a beam consisting mostly of muon neutrinos ($\nu_\mu$), as well as the corresponding anti-neutrino process \cite{MiniBooNE:2018esg, MiniBooNE:2020pnu}. As the baseline (that is, the distance between the neutrino source and the detector) in MiniBooNE is far too short for standard 3-flavor oscillation to develop, this result has led to a flurry of studies investigating the possibility that a fourth, sterile, neutrino is responsible for the excess, either via oscillations or via its decays \cite{Fischer:2019fbw, Gninenko:2009ks, Bertuzzo:2018itn, Dentler:2019dhz, Ballett:2018ynz, deGouvea:2019qre, Abdallah:2020biq, Dutta:2020scq, Datta:2020auq, Abdallah:2020vgg, Abdullahi:2020nyr, Brdar:2020tle, Abdallah:2020biq, Vergani:2021tgc, Babu:2022non}. At the same time, MiniBooNE has also driven significant progress in our understanding of neutrino--nucleus interactions in the Standard Model (SM)~\cite{%
  Hill:2009ek,              
  Martini:2009uj,           
  Nieves:2011pp,            
  Sobczyk:2012ms,           
  Meucci:2012yq,            
  Lalakulich:2012ac,        
  Meloni:2012fq,            
  Nieves:2012yz,            
  Lalakulich:2012hs,        
  Zhang:2012xn,             
  Martini:2012uc,           
  Aguilar-Arevalo:2013pmq,  
  Formaggio:2012cpf,        
  Coloma:2013rqa,           
  Wang:2013wva,             
  Mosel:2013fxa,            
  Wang:2014nat,             
  Megias:2014qva,           
  Ericson:2016yjn,          
  MiniBooNE:2018esg,        
  Ioannisian:2019kse,       
  Giunti:2019sag,           
  Brdar:2021ysi,            
  Alvarez-Ruso:2021dna},    
progress that will also be invaluable for the next generation of neutrino oscillation experiments. However, none of these studies has identified effects that would be large enough to account for the MiniBooNE excess, so at the moment no known SM explanation for the anomaly exists.  The same conclusion has been reached in Ref.~\cite{Brdar:2021ysi}, which critically examined how different Monte Carlo (MC) event generators differ in their background predictions for the MiniBooNE $\nu_e$ appearance search.

In this paper, we will complement the results from Ref.~\cite{Brdar:2021ysi} in several ways. First, we will use the latest MiniBooNE data from Ref.~\cite{MiniBooNE:2020pnu}, while Ref.~\cite{Brdar:2021ysi} was based on the previous data release, Ref.~\cite{MiniBooNE:2018esg}.  More importantly, though, we will introduce in \cref{sec:2p2hFormalism} a new contribution to MiniBooNE's background budget, namely two-particle-two-hole (2p2h) scattering with final-state radiation (``2p2h$\gamma$''). Because 2p2h interactions can be viewed as a neutrino interacting with a pair of tightly bound nucleons rather than a single nucleon, the probability for photon emission can be enhanced by a coherence factor ${\sim}2$.  We will discuss this process in the context of MiniBooNE, but will also present predictions for liquid argon detectors which may be able to identify and reconstruct 2p2h$\gamma$ events.

In addition, in \cref{sec:NCPi0}, we will revisit the NC$\pi^0$ background. These are events in which a single neutral pion (and no other visible particles) is produced, but its decay products are (mis-)reconstructed as a single electromagnetic shower, thus mimicking a charged-current quasi-elastic (CCQE) $\nu_e$ interaction -- the signal that MiniBooNE is looking for. We construct a purely phenomenologically-driven approach to determine this background that matches the full approach of MiniBooNE very well. In applying this approach to other neutrino event generators, we find that other generators tend to favor a larger NC$\pi^0$ rate in the low-energy region where the anomaly is observed, even when data-driven methods are used to fix the overall normalization.

Combining all these effects, we find that the significance of the MiniBooNE anomaly may be slightly smaller than previously estimated, however it still remains tantalizingly high. Throughout this work, we offer some perspective on how current and near-future liquid argon detectors may be able to perform their own independent analyses of these interesting single- and double-photon neutrino-scattering processes.
  
\section{Single{-}Photon Events from 2p2h\texorpdfstring{$\gamma$}{+photon} Scattering}
\label{sec:2p2hFormalism}

In this section, we elaborate on the contribution of 2p2h processes with final state radiation (``2p2h$\gamma$'') to the background budget in MiniBooNE. We also discuss possible dedicated searches for such processes in future detectors, in particular in liquid argon time-projection chambers (TPCs).

\begin{figure}[t]
    \centering
    \includegraphics[width=0.6\linewidth]{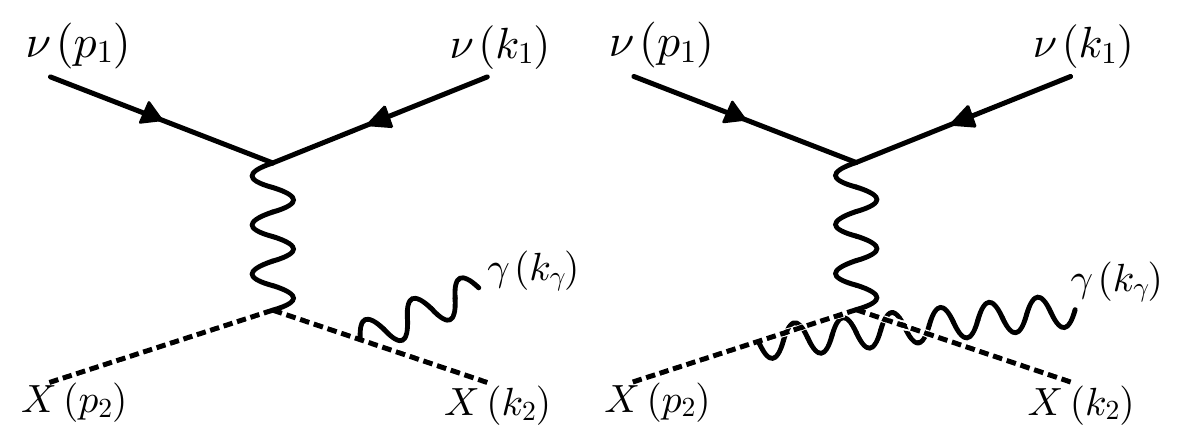}
    \caption{Feynman diagrams for 2p2h$\gamma$ emission in our toy model, in which the process is described as the radiative scattering of a neutrino on a tightly bound 2-nucleon system $X$.}
    \label{fig:FD}
\end{figure}

Our goal here is not to carry out a full-fledged calculation of the cross-section for 2p2h$\gamma$ processes. Instead, we work with a toy model in which we assume that the neutrino scatters off a bound two-particle system $X$, assumed to be a tightly bound pair of protons. (Scattering on neutron pairs can be ignored here because electrically neutral particles cannot emit final state radiation.) We treat $X$ as a scalar particle with mass $m_X = 2 m_N \approx \SI{2}{GeV}$, interacting with the neutrino (and possibly charged lepton) via $Z$/$W$ bosons. \Cref{fig:FD} demonstrates the two Feynman diagrams that contribute to 2p2h$\gamma$ processes in this toy model.

\subsection{Scattering without Photon Emission}
\label{subsec:WithoutPhoton}

We begin by calculating the cross-section for charged-current scattering $\nu_\mu + X \to \mu^- + X'$ in our toy model and comparing the result to more accurate calculations from the literature. This serves as a validation of the toy model. The relevant Feynman diagram is similar to the diagrams shown in \cref{fig:FD}, with the photon removed and with $W$-exchange instead of $Z$-exchange. We assume $m_X \approx m_{X'}$. This gives the matrix-element-squared (in the limit $s \ll m_W^2$)
\begin{align}
    &\left| \mathcal{M} \right|^2 = 32 G_F^2 \left\lvert F(Q^2)\right\rvert^2
      \left[ 4s(s+t) - 4m_X^2(s-m_X^2) - m_\mu^2\left(4s+t-m_\mu^2\right)\right],
\end{align}
where $G_F$ is the Fermi constant, $m_\mu$ is the muon mass, $s$ and $t$ are Mandelstam variables, and $F(Q^2)$ is a form factor that parameterizes the ability of the $W$ boson to resolve the substructure of the two-nucleon system (leading to suppression at high $Q^2$) as well as Pauli-blocking of final state nucleons (leading to suppression at low $Q^2$).

\begin{figure}
    \centering
    \includegraphics[width=0.5\linewidth]{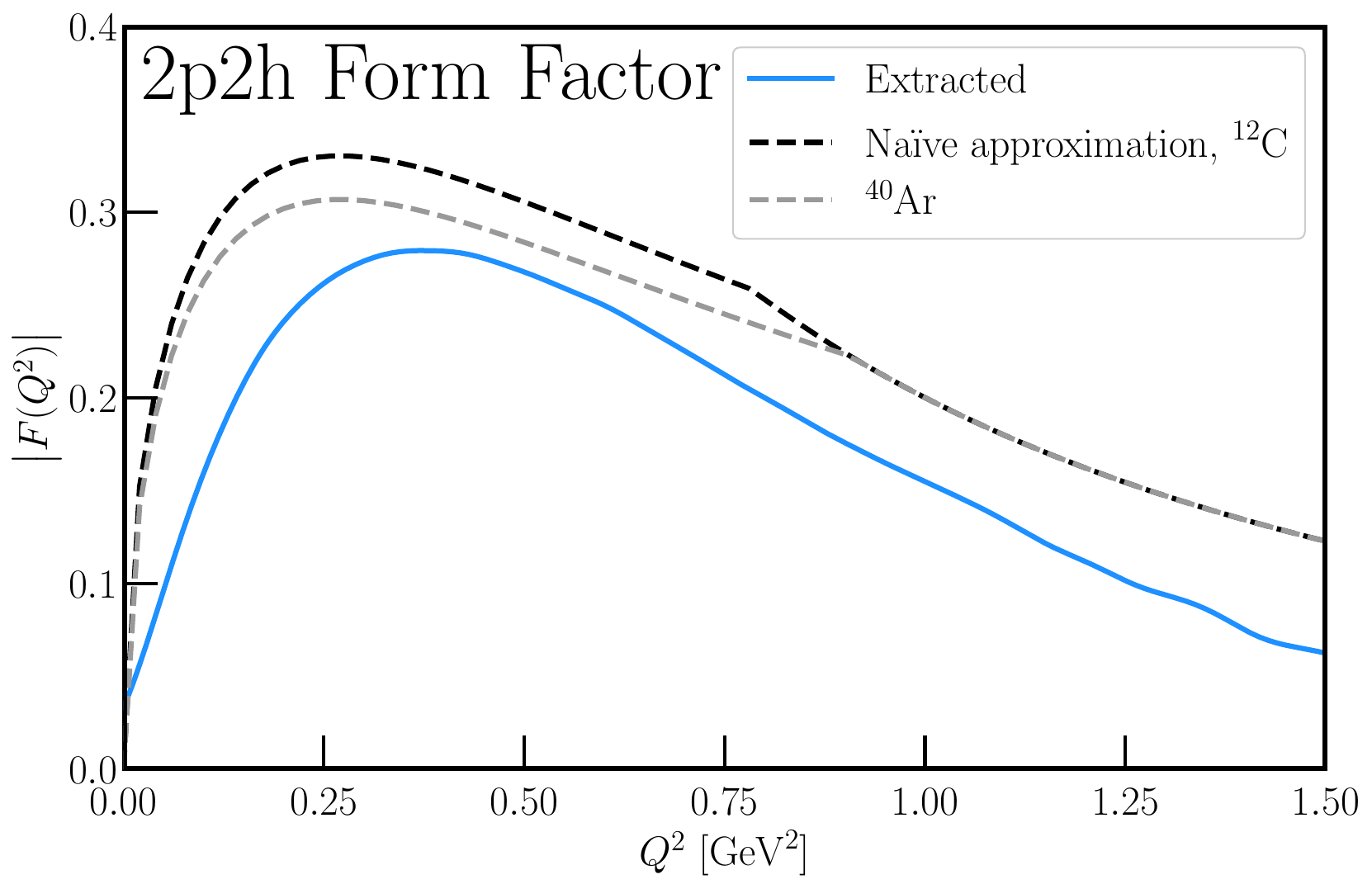}
    \caption{Form factor $|F(Q^2)|$ for two-particle--two-hole scattering of a neutrino off a scalar target $X$ with mass $2m_N$ in our toy model as a function of the momentum-transfer-squared $Q^2$. The blue curve has been derived by matching the differential flux-weighted cross-section on \iso{C}{12} in our toy model against the results of Ref.~\cite{Gallmeister:2016dnq} for the same target nucleus. The dashed black line corresponds to the theoretical estimate for the form factor from \cref{eq:F} for the same isotope. We find good, but not perfect, agreement. The gray curve shows the theoretical estimate for \iso{Ar}{40}.}
    \label{fig:FF}
\end{figure}

In order to compare the predictions of this toy model against existing results, we determine the differential cross-section as a function of $Q^2 \equiv -t$, weighted by the $\nu_\mu$ flux at MiniBooNE~\cite{MiniBooNE:2008hfu}. We first set $F(Q^2) = 1$ in our calculation and then compare the resulting flux-weighted differential cross-section against the results from Ref.~\cite{Gallmeister:2016dnq} to extract an empirical form for $F(Q^2)$. This is shown in \cref{fig:FF} (solid blue line) for $Q^2$ between 0 and \SI{1.5}{GeV^2}, the range of interest for scattering of neutrinos with energy $E_\nu \lesssim \si{GeV}$.  As we extend to the case \text{with} photon emission, we assume that the form factor of the $X$ response to a momentum transfer $Q$ is identical to this case. The extracted form factor includes additional physics on top of the true form factor. For instance, binding-energy effects are relevant for low $Q^2$, where ejection of two nucleons from the nucleus is impossible.  

We can compare the empirically determined $F(Q^2)$ against a simple analytical estimate for validation. To obtain the latter, we envision the scattering off a multi-nucleon system as relevant only for a narrow range of momentum transfers. On the one hand, the outgoing nucleons' momentum must be large enough to overcome Pauli blocking inside the nucleus. But on the other hand, the momentum transfer should be low enough not to resolve the substructure of the multi-nucleon system.  We model the effect of Pauli blocking as a form factor
\begin{align}
  F_P(Q^2) \equiv \min\bigg(\frac{\sqrt{Q^2}}{4 k_F}, 1 \bigg) \,,
  \label{eq:FP}
\end{align}
where $k_F$ is the nucleon Fermi momentum which we estimate as $k_F = [3 \pi^2 A / (4 V)]^{1/3}$, with the nuclear mass number $A$ and the volume of the nucleus, $V$. To compute $V$, we use nuclear charge radii from the table of isotopes in Ref.~\cite{Angeli:2013}.  Note that $V$ is approximately proportional to $A$, so that the dependence of $F_P(Q^2)$ on $A$ drops out approximately. The factor $4$ in the denominator of $k_F$ accounts for the fact that protons and neutrons form independent Fermi gases and that there are two spin states.  As we are interested only in relatively light nuclei ($^{12}$C and $^{40}$Ar) in this paper, we neglect the fact that the proton and neutron Fermi momenta can be slightly different when the numbers of protons and neutrons are not the same.  We model the transition to fully quasi-elastic scattering with a dipole form factor
\begin{align}
  F_D(Q^2) = \bigg(1 + \frac{Q^2}{m_D^2} \bigg)^{-2} \,,
  \label{eq:FD}
\end{align}
where $m_D$ sets the scale of the transition, which we empirically choose to be $m_D = \SI{900}{MeV}$. Overall, we then have
\begin{align}
    F(Q^2) \approx F_P(Q^2) \, F_D(Q^2) \,.
    \label{eq:F}
\end{align}
Besides this form factor, we also incorporate bound-state effects by modeling the two-nucleon system as having a randomly oriented $\sim \SI{200}{MeV}$ initial-state Fermi momentum and a binding energy of $\sim \SI{30}{MeV}$ so that momentum transfers below $Q^2 \approx \SI{0.2}{GeV^2}$ are suppressed. This combination of assumptions yields the dashed black curve in \cref{fig:FF}, not dissimilar from the extracted curve. We will use the extracted curve in the remainder of this work.

\subsection{Scattering with Photon Emission}
\label{subsec:WithPhoton}

The diagrams in \cref{fig:FD} can be calculated to determine the matrix-element-squared with photon emission in our toy model. We provide the full expression for $\left|\mathcal{M}\right|^2$ in \cref{app:Expressions} as a function of dot-products between different particles' four-momenta. Numerically, we work in the rest-frame of the $\nu+X$ scattering, using the direction of the outgoing photon, the photon's energy, and the outgoing $X$ energy as our kinematic quantities. Note that scattering with photon emission is possible only for scattering on pairs of protons, but not for scattering on neutron pairs.

The momentum transfer $Q^2$ can be determined from the outgoing and incoming neutrino four-momenta, $k_1$ and $p_1$, according to $Q^2 = -\left(k_1 - p_1\right)^2$. The full expression for $Q^2$ in terms of the observed energies and momenta, which we use in our numerical calculations, is given in \cref{app:Expressions} as \cref{eq:QsqFull}, but it is illuminating to consider also the momentum transfer averaged over the outgoing photon directions,
\begin{equation}\label{eq:QsqAvg}
    \overline{Q^2} = \frac{\left(s - m_X^2\right)
        \left(\sqrt{s} - E_\gamma^\text{cm} - E_X^\text{cm} \right)}{\sqrt{s}} \,.
\end{equation}
Here, $s = m_X^2 + 2 m_X E_\nu$ is the center-of-mass energy squared, $E_\gamma^\text{cm}$ is the outgoing photon energy, and $E_X^\text{cm}$ is the energy of the outgoing two-nucleon system. The superscript ``cm'' indicates that these quantities are in the center-of-mass frame. In calculating the cross-section for 2p2h$\gamma$ interactions, we use the form factor $F(Q^2)$ extracted for the 2p2h process (solid blue curve in \cref{fig:FF}), assuming the two-nucleon response to a $Z$ boson is the same as that to a $W$ boson. We utilize \textsc{vegas}~\cite{Lepage:2020tgj} to integrate over phase space.

\begin{figure}
    \centering
    \includegraphics[width=0.5\linewidth]{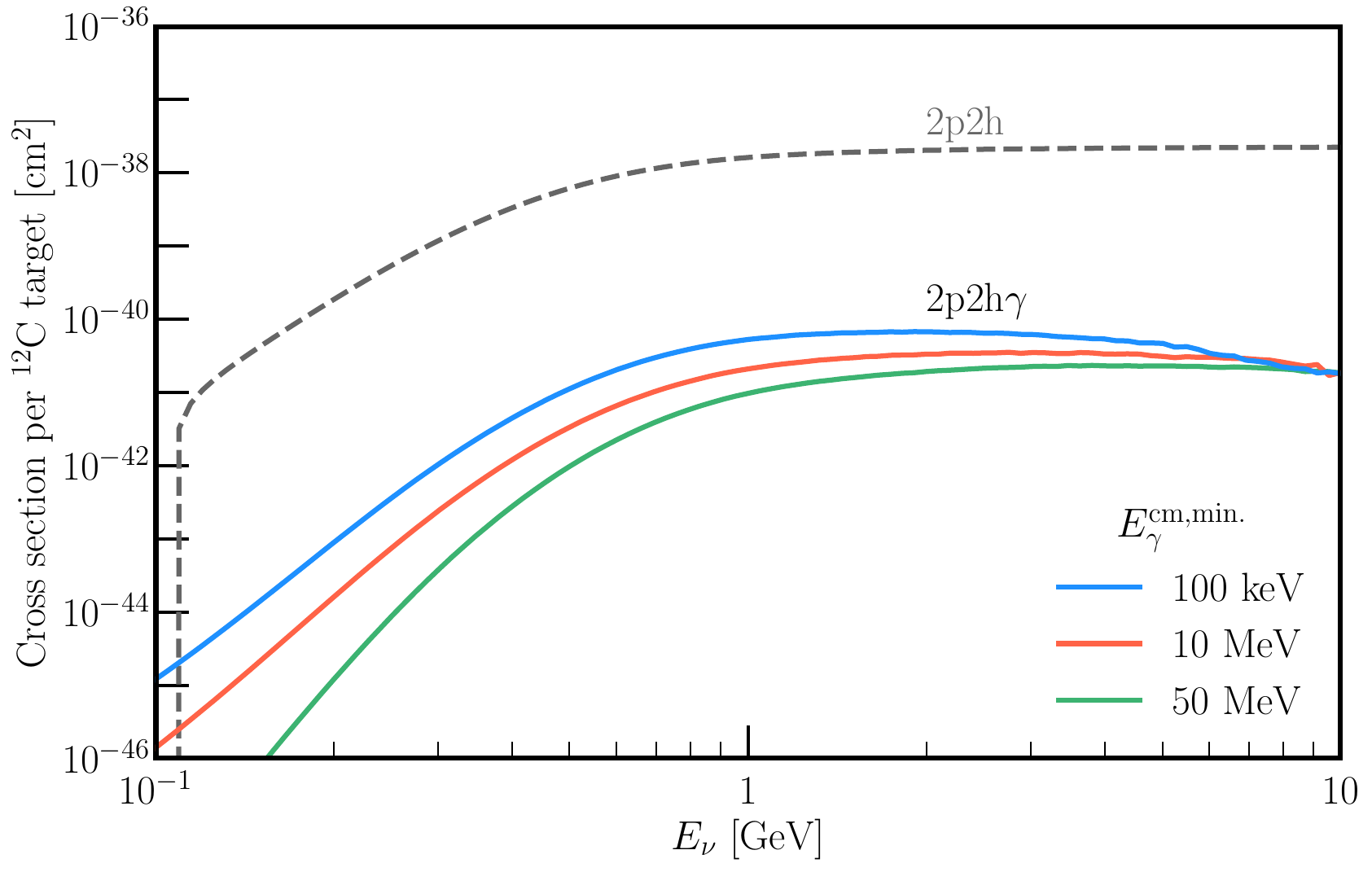}
    \caption{Total cross-section of neutrino scattering off a pair of nucleons, $X$, in our formalism, comparing charged-current scattering $\nu_\mu  X \to \mu^- X$ (dashed) against neutral current scattering with final-state radiation, $\nu_\alpha  X \to \nu_\alpha X \gamma$ (colored). We take three values for the minimum (CoM-frame) outgoing photon energy $E_\gamma^\text{cm,min.}$ -- $\SI{100}{keV}$ (blue), $\SI{10}{MeV}$ (orange), and $\SI{50}{MeV}$ (green). We normalize all cross-sections here to the number of $^{12}$C targets.}
    \label{fig:XSec}
\end{figure}

The resulting cross-section is logarithmically sensitive to the minimum (center-of-mass-frame) photon energy, $E_\gamma^\text{cm,min.}$, used as an infrared cutoff in the calculation. \Cref{fig:XSec} compares this cross-section for neutrino energies between $\SI{100}{MeV}$ and $\SI{10}{GeV}$ for three choices of $E_\gamma^\text{cm,min.}$: $\SI{100}{keV}$ (blue), $\SI{10}{MeV}$ (red), and $\SI{50}{MeV}$ (green). Comparing these against the 2p2h cross-section discussed in \cref{subsec:WithoutPhoton} (dashed black), we find a similar dependence on $E_\nu$, but an overall suppression by about two orders of magnitude, corresponding to the extra factor $\alpha$ (electromagnetic fine structure constant) in the 2p2h$\gamma$ cross-section.

\subsection{2p2h\texorpdfstring{$\gamma$}{+photon} in MiniBooNE}
\label{subsec:MBResults}

Folding the cross-sections obtained in \cref{subsec:WithPhoton} with the neutrino flux of the Booster Neutrino Beam~\cite{MiniBooNE:2008hfu}, we determine the 2p2h$\gamma$ event rate in MiniBooNE. After simulating 2p2h$\gamma$ events, we deliberately misinterpret the outgoing lab-frame photon as an electron from charged-current quasi-elastic (CCQE) $\nu_e$ scattering\footnote{We also include a 10$\degree$ angular uncertainty and a fractional energy uncertainty of $(8\%/\sqrt{E_\gamma \ [\mathrm{GeV}]} \bigoplus 2\%)$ on the outgoing photons in MiniBooNE \cite{MBtalk}. In the last expression, $\bigoplus$ indicates that the two contributions to the uncertainty are statistically uncorrelated and should therefore be added in quadrature.} and, based on this (incorrect) assumption, we determine the would-be reconstructed neutrino energy using~\cite{MiniBooNE:2010bsu, Brdar:2021ysi}
\begin{align}
    E_\nu^{\rm QE} = \frac{2m'_n E_\gamma - (m'^2_n + m_e^2 - m_p^2)}
                             {2\left(m'_n - E_\gamma + \sqrt{E_\gamma^2 - m_e^2}\cos\theta_\gamma\right)} \,,
    \label{eq:EReco}
\end{align}
where $m_e$ and $m_p$ are the electron and proton masses, respectively, and $m'_n = m_n - E_B$ is the neutron mass minus the binding energy. We set $E_B = 0$ in our analyses to be consistent with the results presented in Ref.~\cite{MiniBooNE:2020pnu}. To compare against MiniBooNE's data on the low-energy excess, we only consider events with $E_\nu^{\rm QE} > \SI{200}{MeV}$. In order to determine the efficiency of reconstructing these events, we compare the reconstructed neutrino energy distribution that we obtain when simulating $\nu_e$ CCQE events with those in Ref.~\cite{MiniBooNE:2020pnu}, obtaining efficiencies on the order of 20--30\% (consistent with those used by the MiniBooNE collaboration in their electron-neutrino analyses).

\begin{figure}
    \centering
    \includegraphics[width=0.6\linewidth]{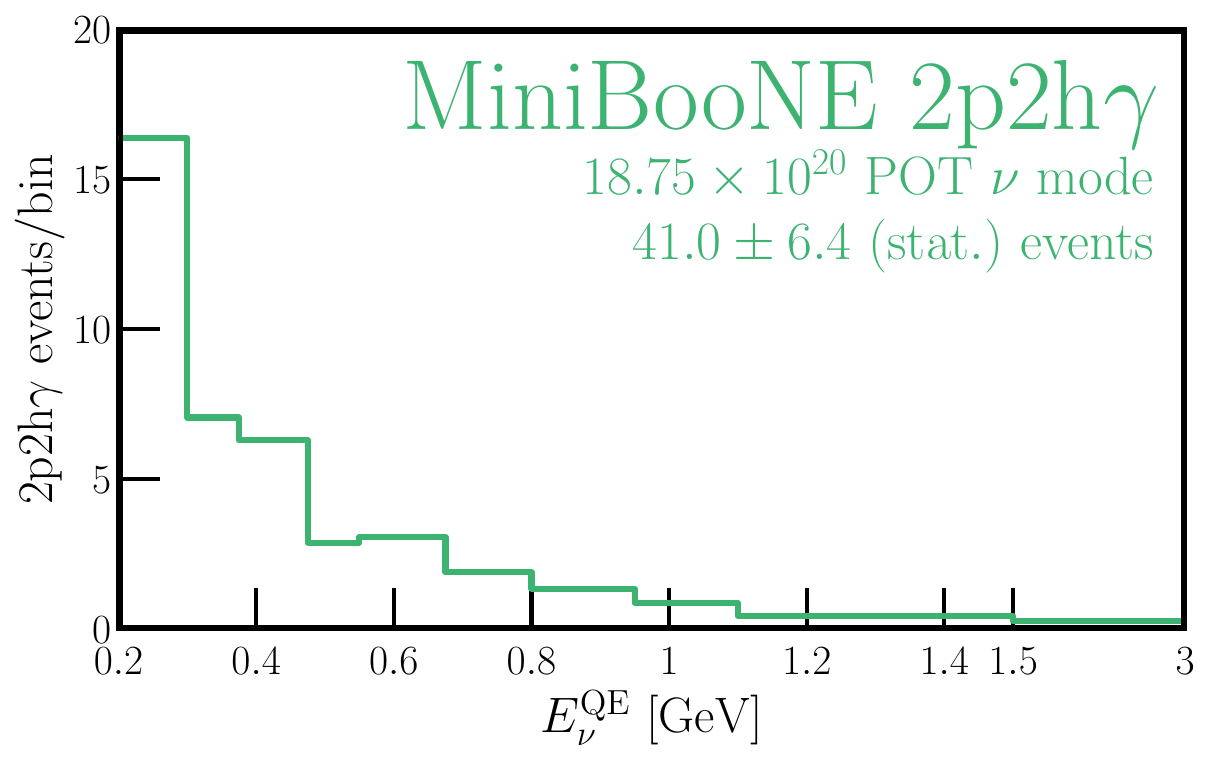}
    \caption{Predicted rate of 2p2h$\gamma$ events in the $18.75 \times 10^{20}$ POT neutrino-mode exposure of MiniBooNE according to our toy model. The total number of predicted events is 41.0.}
    \label{fig:MBEvts_EQE}
\end{figure}

After applying these efficiencies, we obtain our main result for MiniBooNE: the 2p2h$\gamma$ event rate for the $18.75 \times 10^{20}$ proton-on-target exposure (corresponding to the neutrino-mode data presented in Ref.~\cite{MiniBooNE:2020pnu}) is $41.0 \pm 6.4$ events, where the quoted uncertainty is statistical and assumed to be larger than systematic uncertainties. The green histogram in \cref{fig:MBEvts_EQE} presents the distribution of these events as a function of $E_\nu^{\rm QE}$ using the same $E_\nu^\mathrm{QE}$ binning as customarily used by the MiniBooNE collaboration~\cite{MiniBooNE:2020pnu}.

Let us remark that the MiniBooNE collaboration do not predict their single-photon backgrounds from first principles as we do here, but rather use the measured $\pi^0$ production rate for data-driven normalization. However, the translation of this control sample into a prediction for the single-photon signal still requires theory input. If the 2p2h$\gamma$ process is not included, the data-driven single-photon prediction would be biased in the same way as the first-principles one.

While the number of predicted 2p2h$\gamma$ events is a small fraction of the excess observed by MiniBooNE (560.6 neutrino-mode events), incorporating 2p2h$\gamma$ events in the background budget reduces the statistical significance of the neutrino-mode excess from $4.69\sigma \to 4.27\sigma$, a meaningful difference when interpreting results. The expected spectrum of 2p2h$\gamma$ events is shown in \cref{fig:MBEvts_EQE}. In addition to reducing the significance of the excess, we highlight here that the shape of the excess also changes due to the fact that the 2p2h$\gamma$ spectrum peaks at low $E_\nu^{\rm QE}$. This leads to an excess that is less peaked at low $E_\nu^{\rm QE}$ than with MiniBooNE's nominal background model, making it more consistent with the shape predicted by the $3+1$ sterile neutrino hypothesis~\cite{MiniBooNE:2020pnu}.

We conclude our discussion of the MiniBooNE results by summarizing: including 2p2h$\gamma$ events in the background simultaneously (a) reduces the significance of the low-energy excess by $\sim 0.4\sigma$ and (b) can be expected to improve the goodness of a $3+1$ sterile neutrino fit.

\subsection{Predictions for Liquid Argon Detectors}
\label{subsec:2p2hg-LAr}

To end this section on the 2p2h$\gamma$ process, we discuss implications for future experiments, in particular those based on liquid argon time projection chamber (LArTPC) technology. We have in mind in particular the detectors comprising the short-baseline neutrino (SBN) program at Fermilab~\cite{MicroBooNE:2015bmn, Machado:2019oxb}, consisting of SBND~\cite{SBND:2020scp}, MicroBooNE~\cite{MicroBooNE:2016pwy}, and ICARUS~\cite{ICARUS:2004wqc}. As part of testing the MiniBooNE low-energy excess, the SBN detectors will search for electron-like signals, photon-like signals, and more exotic signals (such as di-electrons)~\cite{Bertuzzo:2018itn, Bertuzzo:2018ftf, Ballett:2018ynz, Ballett:2019pyw, Abdullahi:2020nyr, Datta:2020auq, Dutta:2020scq, Abdallah:2020biq, Abdallah:2020vgg, Hammad:2021mpl, Dutta:2021cip}. MicroBooNE has begun this process with its first datasets, observing results consistent with the SM in photon-based~\cite{MicroBooNE:2021zai}\footnote{We remark that the analysis presented in Ref.~\cite{MicroBooNE:2021zai} is optimized for photons from the decays of $\Delta(1232)$ resonances and therefore would not be as sensitive to the 2p2h$\gamma$ contribution we focus on here.} and electron-based~\cite{MicroBooNE:2021ktl, MicroBooNE:2021nxr, MicroBooNE:2021bcu, MicroBooNE:2021pld} searches. As more data are collected (including at SBND and ICARUS), additional analyses will be developed that will allow for searches for more specific final states, including an inclusive $1\gamma0\ell X$ search, similar to the inclusive $1eX$ search presented in Ref.~\cite{MicroBooNE:2021nxr}.

A key feature of these detectors is their ability to distinguish between electrons and photons, unlike MiniBooNE. Moreover, final state protons will typically be visible as well. According to our toy model, the ideal signature to isolate 2p2h$\gamma$ events in a LArTPC would be two protons plus an electromagnetic shower displaced from the primary vertex due to the photon conversion distance (``$2p1\gamma$'' events). In reality, of course, it is possible that one or both of the protons do not leave the nucleus, so 2p2h$\gamma$ interaction will also contribute to $1p1\gamma$ and $0p1\gamma$ events.

To estimate the rate of 2p2h$\gamma$ events in LArTPCs, we follow the same toy formalism as in \cref{subsec:WithPhoton}, but accounting for the fact that an Ar-40 nucleus contains nine proton pairs, compared to just three in C-12. The form factor $F(Q^2)$ from \cref{eq:F} differs only slightly between the two isotopes, given that the dependence on the nuclear mass number $A$ almost cancels in \cref{eq:FP}. (We neglect here the fact that Ar-40 contains slightly more neutrons than protons, whereas C-12 is isospin-symmetric.)  We use a lower energy threshold for LArTPCs compared to MiniBooNE's \v{C}erenkov detector. In fact, MiniBooNE's threshold on the reconstructed neutrino energy, $E_\nu^{\rm reco.} > \SI{200}{MeV}$ (see \cref{eq:EReco}) corresponds to $E_\gamma \gtrsim \SI{100}{MeV}$ in 2p2h$\gamma$ events. Meanwhile, ArgoNeuT has demonstrated that LArTPCs are capable of reconstructing photons down to tens of MeV~\cite{ArgoNeuT:2018tvi}. For the remainder of our discussion of LArTPCs, we consider two reconstruction threshold benchmarks: a conservative threshold $E_\gamma > \SI{30}{MeV}$, and an optimistic one $E_\gamma > \SI{10}{MeV}$. We also account for a conservative $10\degree$ angular resolution \cite{MicroBooNE:2021zai, MicroBooNE:2021nss} and a fractional energy resolution of $15\%/\sqrt{E\ [\mathrm{GeV}]} \bigoplus 2\%$.  This energy resolution is based on Ref.~\cite{DUNE:2015lol} and is consistent with the results of Ref.~\cite{MicroBooNE:2021nss} for measurements of photons coming from $\pi^0$ decays.

\begin{figure*}
\begin{center}
    \includegraphics[width=\linewidth]{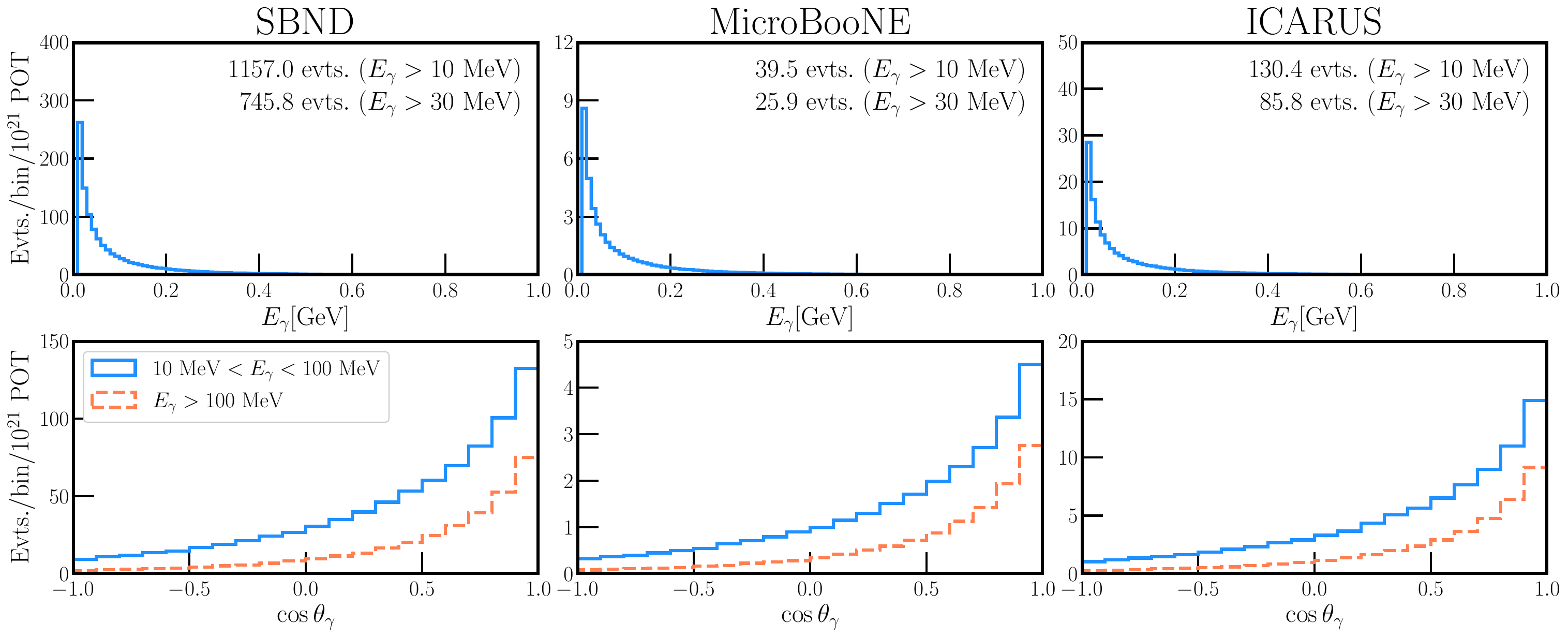}
    \caption{Expected event distributions of 2p2h$\gamma$ single-photon events in the three SBN detectors -- SBND (left), MicroBooNE (center), and ICARUS (right). All distributions are normalized to an exposure of $10^{21}$ protons on target, with the different detector masses and neutrino fluxes at the detector locations factored into each distribution. We have not included any efficiency factors, but we have applied Gaussian energy and angular smearing, as described in the text. Labels on the top row indicate the total number of signal events expected with either a \SI{10}{MeV} or \SI{30}{MeV} photon energy threshold. In the bottom figure, the angular distributions are divided based on those events with low photon energy ($E_\gamma$ between 10 and \SI{100}{MeV}, solid blue) and high photon energy ($E_\gamma > \SI{100}{MeV}$, dashed orange).}
    \label{fig:SBNEvtDists1D}
\end{center}
\end{figure*}

We use the predicted neutrino fluxes and spectra at the three SBN detector locations from Ref.~\cite{MicroBooNE:2015bmn}, and we normalize to an exposure of \SI{e21}{POT}. For this exposure, we find that MicroBooNE can expect to observe $39.5$ ($25.9$) events with $E_\gamma > \SI{10}{MeV}$ (\SI{30}{MeV}). The corresponding number for SBND is $1157.0$ ($745.8$) events thanks to its larger detector mass and shorter baseline. For the even larger but also more distant ICARUS detector it is $85.8$ ($130.4$) events. The predicted photon energy spectrum and angular distribution of these events is shown in \cref{fig:SBNEvtDists1D}. Photons are preferentially emitted in the forward direction ($\cos\theta_\gamma \approx 1$), but the distribution has a long tail reaching all the way to $\cos\theta_\gamma = -1$. Higher energy events tend to be more forward to lower energy ones.

While we have not applied any efficiency factors in \cref{fig:SBNEvtDists1D}, we find it useful to consider some in an attempt to compare against existing MicroBooNE observations. Ref.~\cite{MicroBooNE:2021zai} searched for single-photon events associated with $\Delta(1232) \to N\gamma$ decays, and found that this analysis yields $\sim 4$--$5\%$ reconstruction efficiency, depending on whether or not there is a proton in the final state. With this efficiency, and rescaling for the appropriate number of protons on target, we expect a contribution of $\sim 1$ 2p2h$\gamma$ event in the samples, compared against expectations from other channels of $20.5 \pm 3.65$ events for the single-proton ($1p1\gamma$) final state and $145.1 \pm 13.8$ events for the proton-less ($0p1\gamma$) one. The observed rate of $1p1\gamma$ ($0p1\gamma$) events is $16$ ($153$). Similar to what we found for MiniBooNE in \cref{subsec:MBResults}, we conclude also for MicroBooNE that 2p2h$\gamma$ events can modify the interpretation of the results of Ref.~\cite{MicroBooNE:2021zai} only at the $\lesssim 1\sigma$ level. (This comparison should be taken with a grain of salt given that the efficiencies used here from MicroBooNE's analysis were highly optimized for $\Delta(1232) \to N\gamma$ events. For instance, in the $1p1\gamma$ channel, a cut on the photon--proton invariant mass was imposed.) When projecting forward to SBN and ICARUS, the expectations can change drastically and it may perhaps be possible to demonstrate the existence of 2p2h$\gamma$ events, even though theoretical uncertainties both on our predictions and on the predictions of other processes with similar signatures will still be a challenge.

\section{Double{-}Photon Events from NC\texorpdfstring{$\pi^0$}{ Single Pion} Scattering}
\label{sec:NCPi0}

Now we turn our focus to two-photon backgrounds to the electron-neutrino search at MiniBooNE, focusing specifically on neutral-current single-pion production, $\nu + X \to \nu + \pi^0 + X'$. The $\pi^0$ will decay into a pair of photons, both of which typically convert into electromagnetic showers within the MiniBooNE fiducial volume. However, single-pion events may be mis-reconstructed as events containing a single electromagnetic shower (and therefore indistinguishable from CCQE $\nu_e$ scattering) if
\begin{enumerate}
    \item one of the photons converts outside the fiducial volume,
    \item one of the photons is lost to photo-nuclear absorption before it converts,
    \item the two electromagnetic showers have significant overlap, or
    \item the event is highly asymmetric in the sense that one photons carries much more energy than the other one.
\end{enumerate}

This section is structured as follows: first, in \cref{subsec:CutApproach}, we detail the procedure by which we attempt to reproduce MiniBooNE's NC$\pi^0$ analysis using the \nuance generator. We also speculate on how a data/Monte Carlo disagreement, in terms of the angular uncertainty of the detector, could impact this rate estimate. Then, in \cref{subsec:Generator}, we investigate how predictions for the NC$\pi^0$ background vary when considering MC generators other than \nuance.

\subsection{Cut-based Approach to Reproduce MiniBooNE's Background Predictions}
\label{subsec:CutApproach}

Our first goal is to reproduce MiniBooNE's prediction for the NC$\pi^0$ background from Ref.~\cite{MiniBooNE:2020pnu}. The $\pi^0$/$e^-$ separation in MiniBooNE is driven by a likelihood-based analysis (see Ref.~\cite{MiniBooNE:2018esg}), where the \v{C}erenkov light pattern in each event is fit using both a single-shower (CCQE $\nu_e$ candidate) and double-shower ($\pi^0 \to \gamma\gamma$ candidate) hypothesis, and the final classification of the event depends on which of the two fits yields the larger likelihood. Events classified as single-shower (CCQE-like) represent the quoted NC$\pi^0$ background in Ref.~\cite{MiniBooNE:2020pnu}.  As it is impossible to reproduce this approach without a full detector simulation, we follow a somewhat different strategy which we will describe below.

We begin by simulating general NC neutrino interactions in MiniBooNE and selecting events with at least one $\pi^0$ in the final state. We then sample the $\pi^0 \to \gamma\gamma$ decay~\cite{puig_eschle_phasespace-2019} to obtain events with at least two photons.  We account for the conversion length of $\SI{50}{cm}$ in MiniBooNE's liquid scintillator, ignoring photons that escape the fiducial volume, $R < \SI{5}{m}$, before converting. We also discard events that contain a photon conversion in the veto region $\SI{5}{m} < R < \SI{6.1}{m}$. Finally, we include the effects of photon absorption on nuclei by removing photons that are absorbed before converting. However, we find that this effect is negligible in our analysis.

It is important to keep in mind that in the majority of NC$\pi^0$ events, both photons are reconstructed and events are \emph{correctly} classified by MiniBooNE. Indeed, the MiniBooNE collaboration has published a measurement of the momentum distribution of $\pi^0$ from NC neutrino interactions in their earlier data set~\cite{MiniBooNE:2009dxl}.
\begin{figure}
    \centering
    \includegraphics[width=0.8\linewidth]{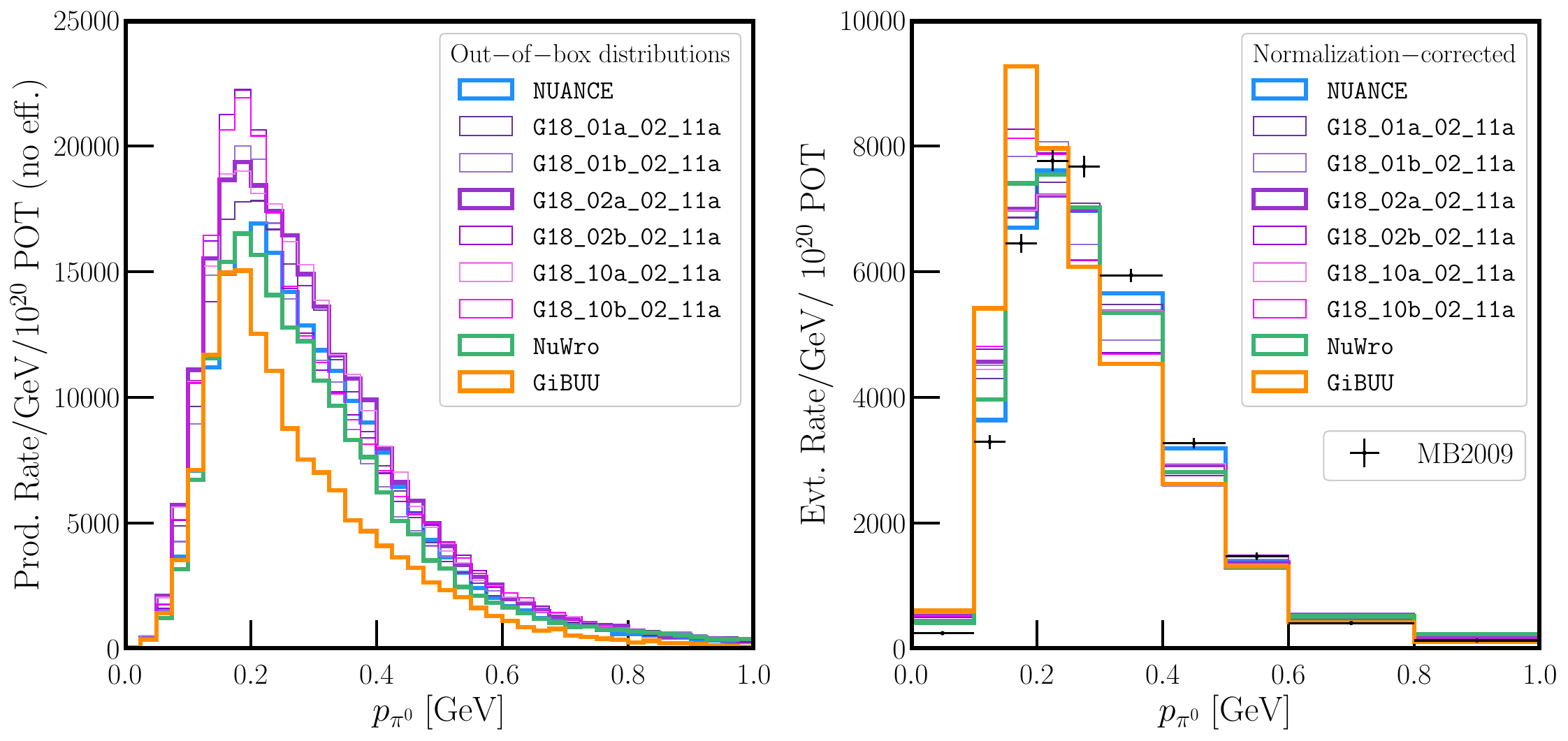}
    \caption{Distribution of reconstructed neutral pion momenta in NC$\pi^0$ events from different neutrino MC generators. The ``out-of-the-box'' distributions shown on the left are obtained directly from the generators before applying MiniBooNE efficiencies, while the ones on the right have been reweighted by the global reweighting factors listed in \cref{tab:Reweight}, as well as MiniBooNE efficiencies, to match the total NC$\pi^0$ production rate measured in Ref.~\cite{MiniBooNE:2009dxl}.}
    \label{fig:PionNorm}
\end{figure}
In order to accommodate this measurement, we re-normalize our sample of Monte Carlo events to match the total production rate of $\pi^0$ measured in Ref.~\cite{MiniBooNE:2009dxl} (but still taking the $\pi^0$ momentum distribution from the generator prediction).\footnote{We have also attempted to re-weight MC events such that they match also the measured $\pi^0$ momentum spectrum. After doing so, the predicted $\pi^0$ background is nearly independent of the event generator used as the only potential remaining difference between generators is then the angular distribution of the $\pi^0$, which has a subleading effect).  The reader may wonder why we do not use this seemingly even more generator-independent approach in our baseline analysis pipeline. The reason is that it is not truly generator-independent either: the signal efficiencies of the analysis from Ref.~\cite{MiniBooNE:2009dxl}, which need to be unfolded in order to obtain the true $\pi^0$ production rate, still depend on Monte Carlo simulations. We therefore choose to use only the overall normalization from Ref.~\cite{MiniBooNE:2009dxl}, but not the shape of the $\pi^0$ spectrum.} This measurement results in an effective reweighting of the out-of-the-box generator samples by the factors given in \cref{tab:Reweight}. The $\pi^0$ distributions before (left) and after (right) reweighting, are shown in \cref{fig:PionNorm} for the different generators that we use in this work. In the following, we always work with re-weighted distributions, but in \cref{app:OutOfBox}, we present also results using the out-of-the-box generator distributions instead of the data-constrained ones.

\begin{table}
    \centering
    \caption{Data-driven reweighting factors for NC$\pi^0$ events in MiniBooNE from different Monte Carlo event generators.}
    \label{tab:Reweight}
    \begin{minipage}{7cm}
        \def\arraystretch{1.2}
        \begin{ruledtabular}
        \begin{tabular}{lr}
            Generator & Reweight Factor \\
            \hline
            \nuance & $1.32$ \\
            \hline
            \texttt{GENIE} G18\_01a\_02\_11a & $1.22$ \\
            \texttt{GENIE} G18\_01b\_02\_11a & $1.26$ \\
            \texttt{GENIE} G18\_02a\_02\_11a & $1.14$ \\
            \texttt{GENIE} G18\_02b\_02\_11a & $1.18$ \\
            \texttt{GENIE} G18\_10a\_02\_11a & $1.14$ \\
            \texttt{GENIE} G18\_10b\_02\_11a & $1.18$ \\
            \hline
            \texttt{NuWro} & $1.44$ \\
            \hline
            \texttt{GiBUU} & $1.91$ \\
        \end{tabular}
        \end{ruledtabular}
    \end{minipage}
\end{table}

We endeavor to approximately reproduce MiniBooNE's efficiency for distinguishing $\pi^0$ events from CCQE $\nu_e$ interactions (characterized by a single electron) in a variety of ways.  We first work with simulated events from the \nuance Monte Carlo generator (the same used by MiniBooNE), deferring to \cref{subsec:Generator} a discussion of how our results depend on the choice of event generator.) We assume the relevant kinematic parameters for the $\pi^0$/$e^-$ separation are the following:
\begin{itemize}
    \item $E_{\rm vis.}$, the total visible energy in the electromagnetic shower(s).
    \item $\cos\theta_{\gamma\gamma}$, the opening angle between the two highest-energy photons in events with at least one $\pi^0$. If only one photon converts in the fiducial volume, then $\cos\theta_{\gamma\gamma}$ is set to $1$ in practice.
    \item $E_{\rm max.}/E_{\rm vis.}$, the fraction of the visible energy carried by highest-energy photon, representing the asymmetry of the shower. Similar to the above, if only one photon converts then this asymmetry is set to $1$ in our simulations.
\end{itemize}
For events with a single $\pi^0$ (the vast majority of NC$\pi^0$ events), these three variables fully describe the kinematics of the $\pi^0$ -- its angle with respect to the beam is not used in our electron/pion discrimination, but impacts results in how it enters the reconstructed neutrino energy $E_{\nu}^{\rm QE}$. Based on the reasoning at the beginning of \cref{sec:NCPi0}, for a given $E_{\rm vis.}$, we expect that NC$\pi^0$ events with $\{\cos\theta_{\gamma\gamma}, E_{\rm max.}/E_{\rm vis.}\}$ near $\{1,1\}$ will tend to pass the likelihood cut and contribute to the background in the $\nu_e$ appearance search in which the MiniBooNE anomaly is manifest.

\begin{figure}
    \centering
    \includegraphics[width=\linewidth]{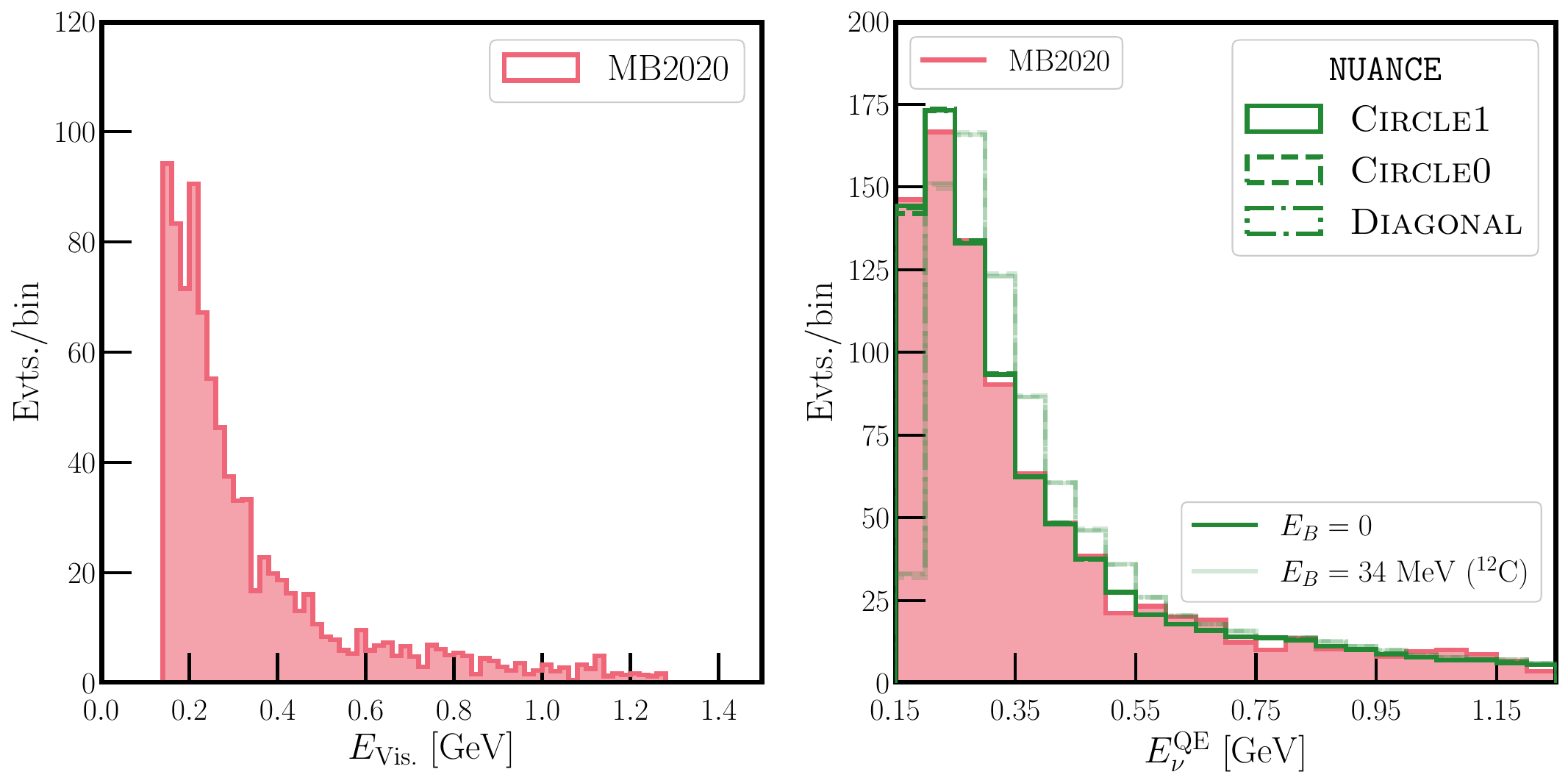}
    \caption{\emph{Left}: Event spectrum in terms of total visible energy of photons in NC$\pi^0$ events mis-reconstructed as CCQE $\nu_e$ interactions. While similar distributions have been shown in Ref.~\cite{MiniBooNE:2020pnu}, the one presented here has been obtained directly from MiniBooNE's sample of NC$\pi^0$ Monte Carlo events, which the collaboration has kindly provided to us.  This allows for finer binning and avoids inconsistencies stemming from the fact that the distributions shown in Ref.~\cite{MiniBooNE:2020pnu} include a cut on $E_{\nu}^{\rm QE}$.\\
    \emph{Right:} Distribution of reconstructed neutrino energy $E_{\nu}^{\rm QE}$ for NC$\pi^0$ events obtained via the method described in \cref{subsec:CutApproach} with the \nuance MC generator and three different cut prescriptions as listed in the legend, compared against the results presented in Ref.~\cite{MiniBooNE:2020pnu} (red histogram). Dark (faint) lines assume $E_B = 0$ ($34$ MeV) in determining the reconstructed energy.}
    \label{fig:EVis_EQE_Validation}
\end{figure}

\begin{figure}
    \centering
    \includegraphics[width=0.5\linewidth]{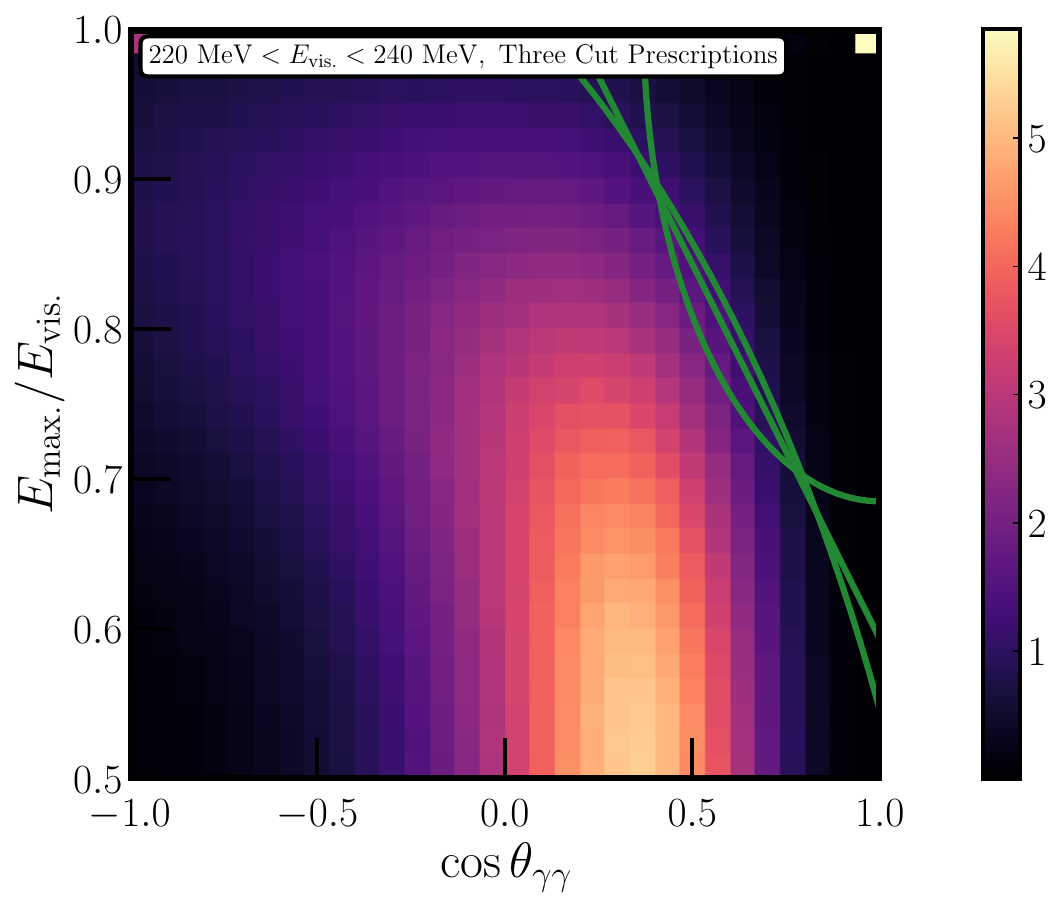}
    \caption{Distribution in $\cos\theta_{\gamma\gamma}$ and $E_{\rm max.}/E_{\rm vis.}$ of \nuance NC$\pi^0$ events with visible energy between \SI{220}{MeV} and \SI{240}{MeV}. The three green lines correspond to the three different types of cuts discussed in \cref{subsec:CutApproach}, determined by requiring that the number of events in this $E_{\rm vis.}$ bin from our full sample of \nuance events matches MiniBooNE's prediction in the same $E_{\rm vis.}$ bin. The bright dot in the upper right-hand corner is from events in which one photon has been absorbed or exits the detector before converting.}
    \label{fig:CutShapesMC}
\end{figure}

Motivated by this, we construct the likelihood using three different empirical methods.  For each of them, we sort events by a parameter $r(\cos\theta_{\gamma\gamma}, E_{\rm max.}/E_{\rm vis.})$ and then impose an $E_{\rm vis.}$-dependent cut, $r < r_{\textsc{cut}}(E_{\rm vis.})$, that is chosen in order to reproduce the $E_{\rm vis.}$ distribution of MiniBooNE's sample of NC$\pi^0$ background Monte Carlo events shown in \cref{fig:EVis_EQE_Validation} (left).  The three different prescriptions for $r$ are the following (see \cref{fig:CutShapesMC} for an illustration):\footnote{We have explored a number of other cuts in the $\cos\theta_{\gamma\gamma}$-vs.-$E_{\rm max.}/E_{\rm vis.}$ plane, all yielding qualitatively similar results to the ones shown here. This includes a cut only on $\cos\theta_{\gamma\gamma}$, similar to the analysis of Ref.~\cite{Brdar:2021ysi}. However, we find that this approach yields significantly larger Monte Carlo uncertainties than the others and therefore do not include it here.}
\begin{itemize}
    \item \textsc{Circle1}: cut along a circle centered at $\{1, 1\}$ in the $\{\cos\theta_{\gamma\gamma}, E_{\rm max.}/E_{\rm vis.}\}$ plane. Events inside the circle are assumed to be mis-reconstructed as CCQE $\nu_e$ interactions. The cut parameter is
    \begin{align}
        r_{\textsc{Circle1}}^2
            = \left(\frac{1 - \cos\theta_{\gamma\gamma}}{2}\right)^2
            + \left(1 - \frac{E_{\rm max.}}{E_{\rm vis.}}\right)^2 \,.
    \end{align}
    \item \textsc{Circle0}: cut along a circle centered at $\{-1, 0\}$ in the $\{\cos\theta_{\gamma\gamma}, E_{\rm max.}/E_{\rm vis.}\}$ plane. Events outside the circle are assumed to be mis-reconstructed as CCQE $\nu_e$ interactions.  The cut parameter is
    \begin{align}
        r_{\textsc{Circle0}}^2
            = \left(\frac{1 + \cos\theta_{\gamma\gamma}}{2}\right)^2
            + \left(\frac{E_{\rm max.}}{E_{\rm vis.}}\right)^2 \,.
    \end{align}
    \item \textsc{Diagonal}: cut along a diagonal in the $\{\cos\theta_{\gamma\gamma}, E_{\rm max.}/E_{\rm vis.}\}$ plane. The cut parameter is
    \begin{align}
        r_{\textsc{Diagonal}}
            = 1 - \frac{1}{2}\left( \frac{1 + \cos\theta_{\gamma\gamma}}{2}
                                  + \frac{E_{\rm max.}}{E_{\rm vis.}} \right) \,.
    \end{align}
\end{itemize}
\cref{fig:CutShapesMC} demonstrates the event distribution as a function of $\cos\theta_{\gamma\gamma}$ and $E_{\rm max.}/E_{\rm vis.}$ for events with $E_{\rm vis.}$ between $\SI{220}{MeV}$ and $\SI{240}{MeV}$, along with curves corresponding to the cut values of $r_{\textsc{Circle1}}$, $r_{\textsc{Circle0}}$, and $r_{\textsc{Diagonal}}$ that allow us to match the expected rate of accepted events by MiniBooNE in that visible energy range. Events to the right and above the cuts are the ones mis-identified as CC $\nu_e$ interactions. We note already here that the event distributions are rising sharply in the part of the parameter space in which the cuts lie, implying that the final distributions of mis-reconstructed events depend sensitively on the shape of these event distributions. Cut values for these three prescriptions for each range of $E_{\rm vis.}$ can be found in \cref{app:CutTable}.

After deriving the above cuts, we are equipped to perform detailed comparisons against various NC$\pi^0$ kinematic distributions presented in Ref.~\cite{MiniBooNE:2020pnu}.  We start in \cref{fig:EVis_EQE_Validation} (right) with the $E_{\nu}^{\rm QE}$ distribution of NC$\pi^0$ events mis-reconstructed as CC $\nu_e$, using uniform bins of $\SI{50}{MeV}$ width (instead of the uneven bin sizes more commonly seen in MiniBooNE plots). We see that all three cut methods (represented by different line styles) reproduce the $E_{\nu}^{\rm QE}$ distribution predicted by the MiniBooNE collaboration (red histogram) very well. However, we only achieve this agreement when setting the binding energy $E_B$ in \cref{eq:EReco} to zero instead of its baseline value for $^{12}$C, \SI{34}{MeV}. Histograms calculated with $E_B = \SI{34}{MeV}$ are shown in fainter colors in \cref{fig:EVis_EQE_Validation} (right). They are similar in shape to the ones with $E_B = 0$, but shifted by about one bin. Indeed, we have confirmed with the MiniBooNE collaboration~\cite{PrivateComms} that $E_B = 0$ has been used when plotting $E_{\nu}^{\rm QE}$ distributions (even though the collaboration's sterile neutrino fits use $E_B = \SI{34}{MeV}$).  Going forward, we will use $E_{B} = 0$ unless otherwise noted.

\begin{figure}
    \centering
    \includegraphics[width=0.85\linewidth]{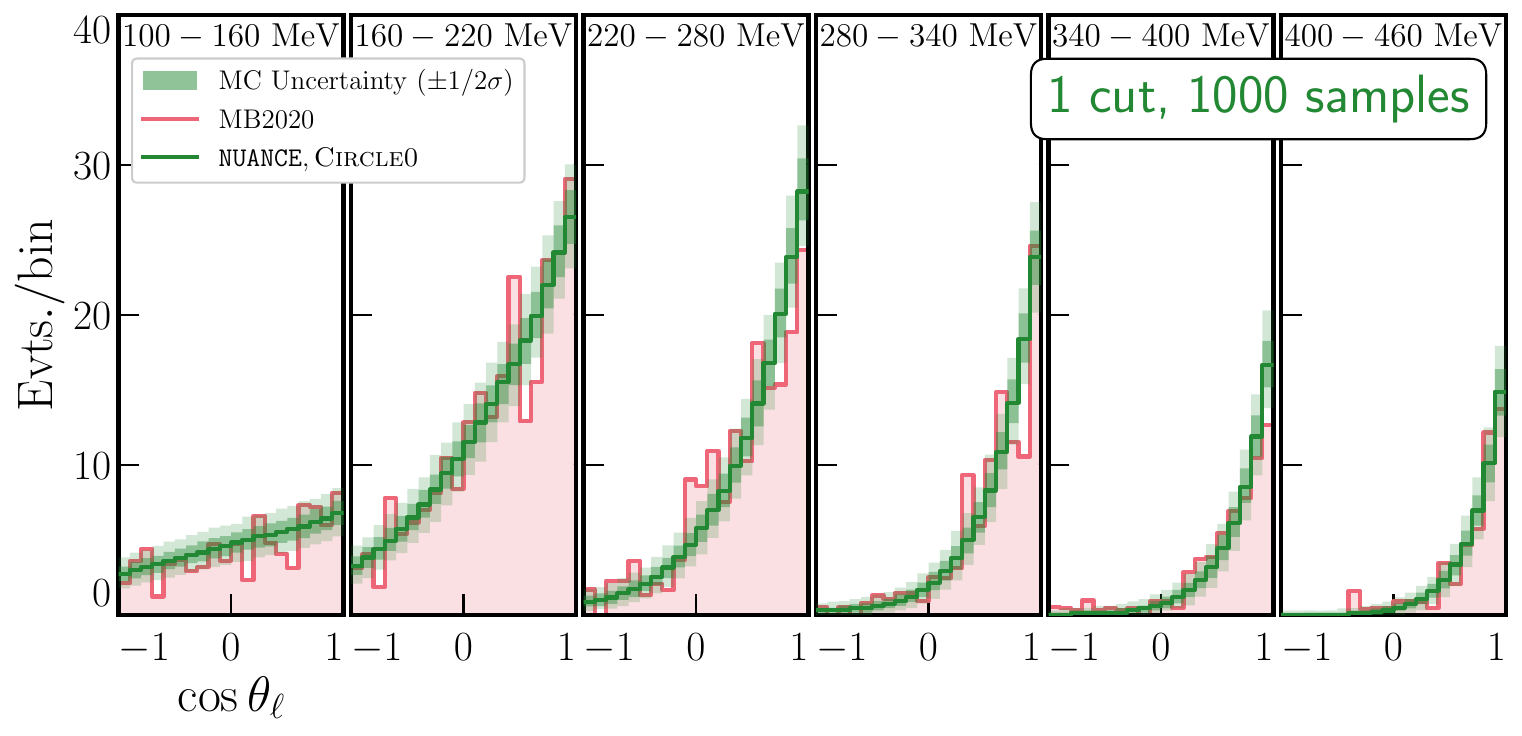}
    \caption{Distributions of NC$\pi^0$ background events in MiniBooNE obtained with the \nuance generator and the \textsc{Circle0} cut, as a function of total visible energy, $E_{\rm vis.}$, and of the direction of the highest-energy photon in the shower with respect to the beam axis, $\theta_\ell$ (thick green lines). Each panel corresponds to a different $E_{\rm vis.}$ range, as labelled at the top. We apply the derived \textsc{Circle0} cuts to 1000 different \nuance subsamples, each similar in size to MiniBooNE's MC sample, and display the $1\sigma$ and $2\sigma$ uncertainty on the event rates from this MC process. The red histogram in the background of each panel presents MiniBooNE's official background prediction from Ref.~\cite{MiniBooNE:2020pnu}.}
    \label{fig:MCSlices}
\end{figure}

\begin{figure}
    \centering
    \includegraphics[width=0.5\linewidth]{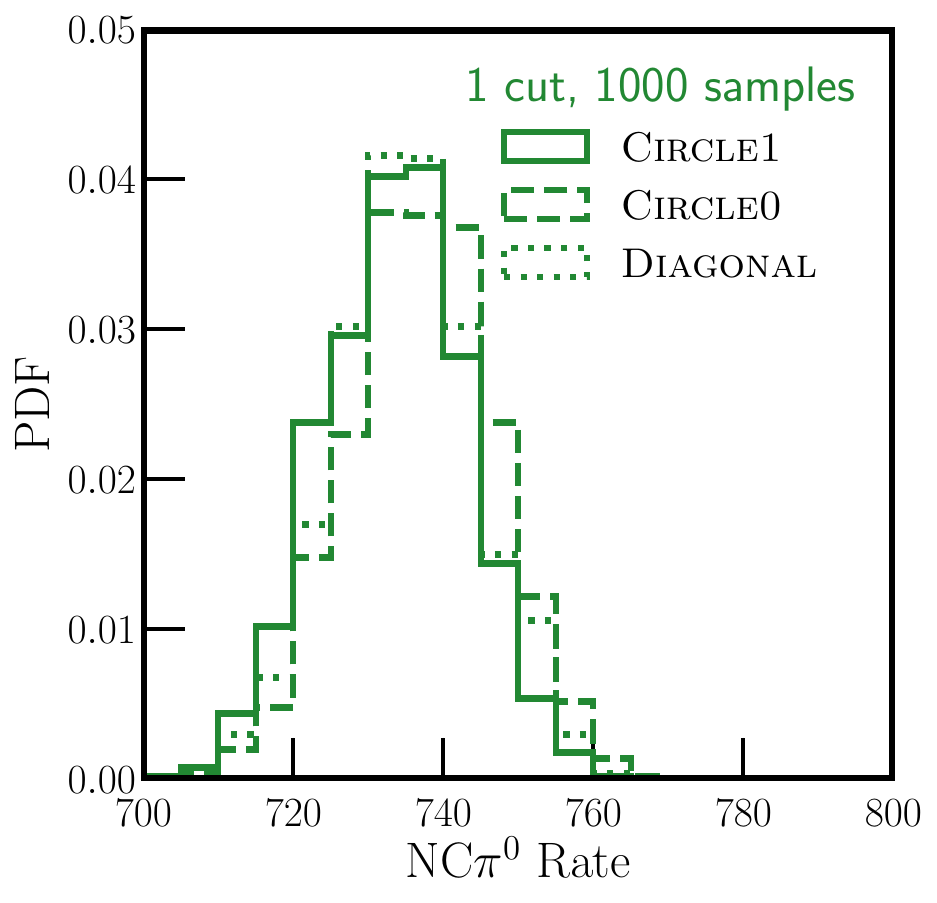}
    \caption{Number of predicted NC$\pi^0$ events in MiniBooNE for different Monte Carlo realizations, each of them similar in statistics to MiniBooNE's. The $e^-/\pi^0$ separation cut has been derived based on our full MC sample and has then be applied to 1000 subsamples.}
    \label{fig:Rates_OneVersion}
\end{figure}

Another distribution of interest for comparison is the angular distributions of would-be electrons (that is, for NC$\pi^0$ background events, the angular distributions of the highest-energy electromagnetic shower) for various slices of $E_{\rm vis.}$. We present our comparison for this distribution (obtained using the \textsc{Circle0} cut method) in \cref{fig:MCSlices}. We have generated 1000 \nuance samples with comparable Monte Carlo statistics to those used in MiniBooNE's determinations of this background~\cite{MiniBooNE:2020pnu}, and show the expected MC uncertainty inferred in this process as colored bands (for $\pm 1\sigma$ and $2\sigma$ ranges).\footnote{It is important to note that Monte Carlo events are weighted, therefore the total number of generated events cannot be used directly to estimate the Monte Carlo statistical uncertainty. Rather, one needs to consider how many of the weighted events lie in the tails of the phase space distributions where they are prone to mis-reconstruction as $e^-$. The bin-to-bin jitter of the histograms in \cref{fig:MCSlices} is a good proxy for this final uncertainty.} This can also be presented in terms of the overall number of events that pass the cuts we have derived for each of the different MC subsamples, which we present in Fig.~\ref{fig:Rates_OneVersion}.
We find that our rates here agree very well with the level of MC statistical uncertainty ($\approx$ 19 events) considered in MiniBooNE's error budget for this background~\cite{MiniBooNE:2020pnu}.

As a side remark, note that our phenomenological approach to determine MiniBooNE's ability to separate pion-like and electron-like events can also be applied in the context of new-physics explanations to the MiniBooNE LEE. For instance, scenarios which posit new particles decaying into $e^+ e^-$ pairs in MiniBooNE's detector~\cite{Bertuzzo:2018itn,Ballett:2018ynz} often lead to overlapping and/or asymmetric electromagnetic showers. The cuts derived here, which depend only on those shower kinematics, can be applied rapidly to estimate the efficiency with which these beyond-the-Standard-Model processes could contribute to the MiniBooNE LEE.

\begin{figure}
    \centering
    \includegraphics[width=0.85\linewidth]{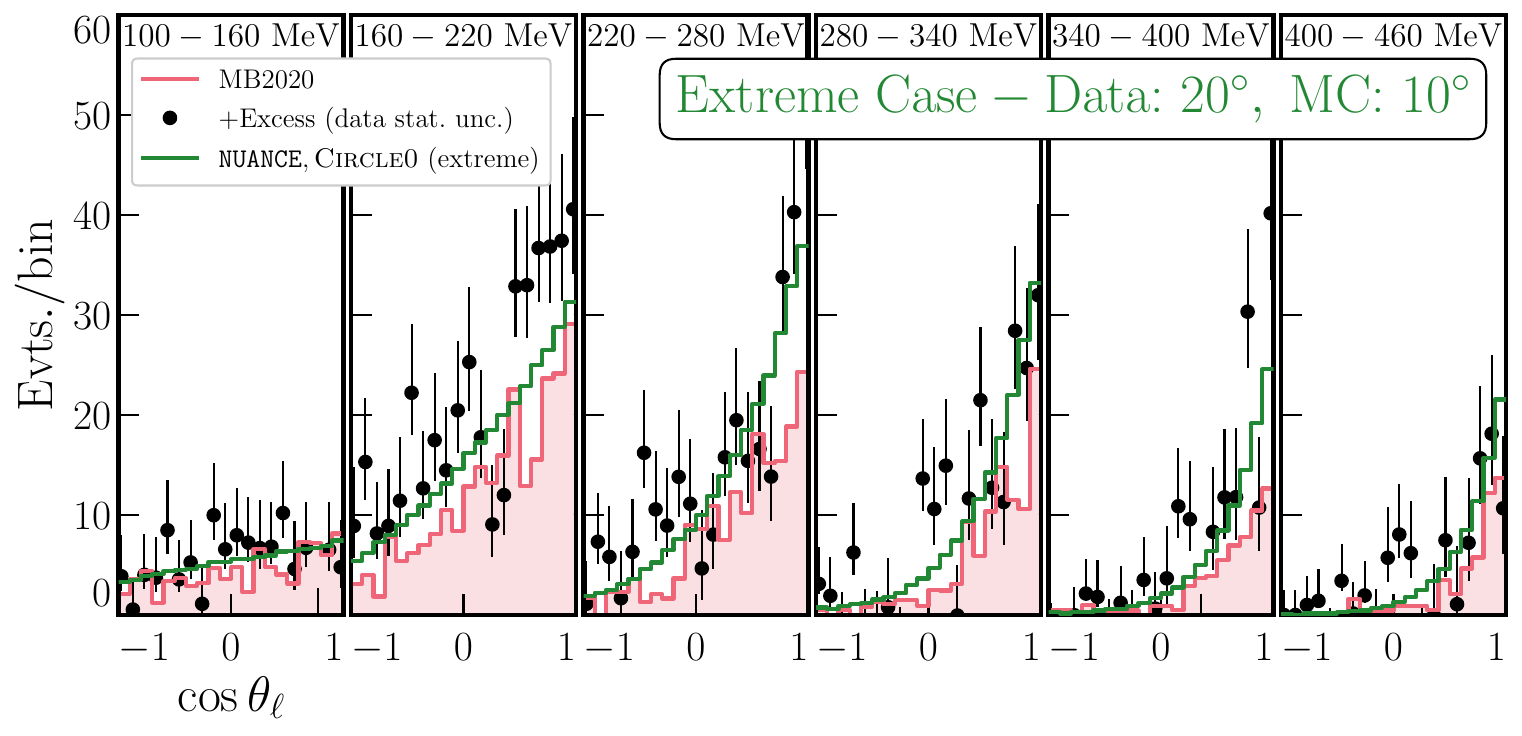}
    \caption{NC$\pi^0$ background events expected in MiniBooNE in an extreme scenario, in which the data/MC disagreement is significantly large: MC events are generated assuming a 10$^\circ$ angular uncertainty, but simulated data assume $20^\circ$. We superimpose the MiniBooNE event excess (with statistical uncertainty according to the data collected) for comparison.  We stress again that the large mismatch between the angular resolution in the simulation compared to the data which we have assumed here serves only to make the resulting bias in the number of events passing cuts clearly visible. We are \emph{not} insinuating that this could explain the MiniBooNE excess, given that MiniBooNE is a very well understood detector.}
    \label{fig:MCSlices_Extreme}
\end{figure}

Our \nuance samples above are all generated assuming the same $10^\circ$ photon angular uncertainty, both when deriving the electron/pion separation cuts, as well as when determining which events pass these cuts, effectively as ``simulated data'' in \cref{fig:MCSlices} and \cref{fig:Rates_OneVersion}. Such a method assumes that the MC and data are well calibrated and that the detector's directional reconstruction abilities are well known. As an \textit{extreme} example, we consider the case that MC simulations are performed (and electron/pion separation determined) with a $10^\circ$ angular uncertainty, but that the collected data actually exhibits a $20^\circ$ uncertainty. This is, of course, an unrealistically large discrepancy for a well-understood detector such as MiniBooNE. The resulting events passing cuts are shown in \cref{fig:MCSlices_Extreme}.
Such a data/MC discrepancy could account for ${\sim}320$ additional NC$\pi^0$ events passing the electron/pion separation cuts and appearing in the low-energy $\nu_e$ search. \cref{fig:MCSlices_Extreme} includes the MiniBooNE low-energy excess in addition to the NC$\pi^0$ background for comparison (with error bars according to the data statistical uncertainty) -- we see that in this extreme scenario, the additional NC$\pi^0$ events passing cuts share similar characteristics to the low-energy excess.

The data/MC discrepancy that we have injected here is, of course extreme, but demonstrative of how such a difference could lead to an enhanced event rate. We leave further exploration of this effect to future work and turn to another difference that could lead to distinct NC$\pi^0$ predictions: those coming from different Monte Carlo neutrino event generators.

\subsection{Comparison against other Monte Carlo Generators}
\label{subsec:Generator}

Given the difference in $p_{\pi^0}$ distributions from different MC generators which we have seen in \cref{fig:PionNorm}, we now turn to the question of whether the event generator used for predicting the NC$\pi^0$ background has an impact on the significance of the MiniBooNE low-energy excess. This question was previously addressed in Ref.~\cite{Brdar:2021ysi}, which concluded that the substitution of one generator for another can reduce the significance by ${\sim}0.5$--$1\sigma$ even when normalizing the MC sample to the measured $\pi^0$ production rate as a function of pion energy.

\begin{figure}
    \centering
    \includegraphics[width=\textwidth]{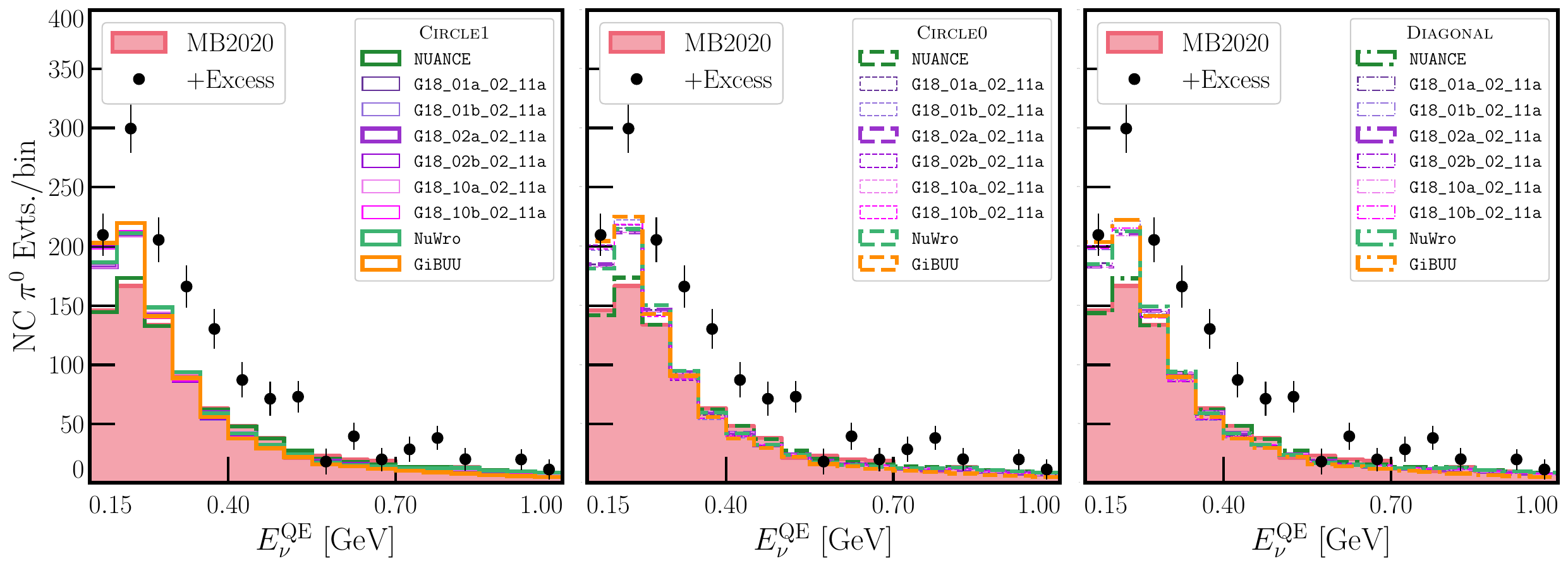}
    \caption{Predicted distributions of NC$\pi^0$ background events in MiniBooNE (neutrino mode) as a function of the reconstructed neutrino energy, $E_{\nu}^{\rm QE}$, from different Monte Carlo generators. In each case, we derive the $\pi^0$/$e^-$ separation cut using our \nuance (the generator used by MiniBooNE) MC samples and apply them to the various generators. The three panels correspond to the three different cut shapes introduced in \cref{subsec:CutApproach}. The official MiniBooNE background prediction~\cite{MiniBooNE:2020pnu} is shown in red.}
    \label{fig:GeneratorComp}
\end{figure}

In this section, we re-consider this question using the more advanced modeling of the $\pi^0$/$e^-$ separation cuts developed in \cref{subsec:CutApproach}. We apply these cuts to MC samples from the generators listed in \cref{tab:Reweight}, namely \nuance
v3.000 \cite{Casper:2002sd}, \texttt{GENIE} v3.00.04 \cite{Andreopoulos:2015wxa}, \texttt{NuWro} v19.02.2-35-g03c3382 \cite{Golan:2012wx},
and \texttt{GiBUU} (2019 release) \cite{Leitner:2008ue}. For \nuance, we use the same input parameters as the MiniBooNE collaboration (input file \texttt{nuance\_defaults\_may07.cards}, flux \texttt{april07\_baseline\_rgen610.6\_flux\_8gev.hbook}). For \texttt{GENIE}, we consider several different tunes, which are explained in more detail in Ref.~\cite{tune-list} (see also the summary in Ref.~\cite{Brdar:2020tle}). In the naming convention \texttt{G18\_XXy\_02\_11a}, \texttt{XX}=\texttt{01} stands for GENIE's baseline tune, tunes with \texttt{XX}=\texttt{02} feature updated models of coherent and resonant scattering, and \texttt{XX}=\texttt{10} indicates theory-driven tunes. The letter \texttt{y}=\texttt{a},\texttt{b} indicates two different implementations of final state interactions.  The raw MC samples used in this work are for the most part identical to the ones used in Ref.~\cite{Brdar:2021ysi}.  Here, we normalize all of them with the reweighting factors from \cref{tab:Reweight}. The resulting $E_{\nu}^{\rm QE}$ distributions are shown in \cref{fig:GeneratorComp}, presenting a finer binning/slightly wider range of this variable than \cref{fig:EVis_EQE_Validation} (right).

Each panel in \cref{fig:GeneratorComp} corresponds to one of the three cut strategies from \cref{subsec:CutApproach}. In contrast with the \nuance curves (green) in each panel, we find that every other generator considered -- \texttt{GENIE}, \texttt{NuWro}, and \texttt{GiBUU} -- prefers a larger rate of NC$\pi^0$ events at low $E_{\nu}^{\rm QE}$ where the MiniBooNE LEE is most prevalent. However, these generators only tend to predict ${\sim}$30 events more than \nuance in the energy range of interest ($\SI{200}{MeV} < E_{\nu}^{\rm QE} < \SI{1250}{MeV}$), because the data-deriven normalization that we impose leads to moderately fewer events at larger energies. Nevertheless the fact that, once events are normalized, \nuance is the only to predict such small rates at low energies is a curious takeaway. For completeness, in \cref{app:OutOfBox}, we repeat this generator comparison \textit{without} this data-driven normalization.

\subsection{Prospects for Liquid Argon Detectors}
\label{subsec:pi0-LAr}

\begin{figure}
    \centering
    \includegraphics[width=0.5\textwidth]{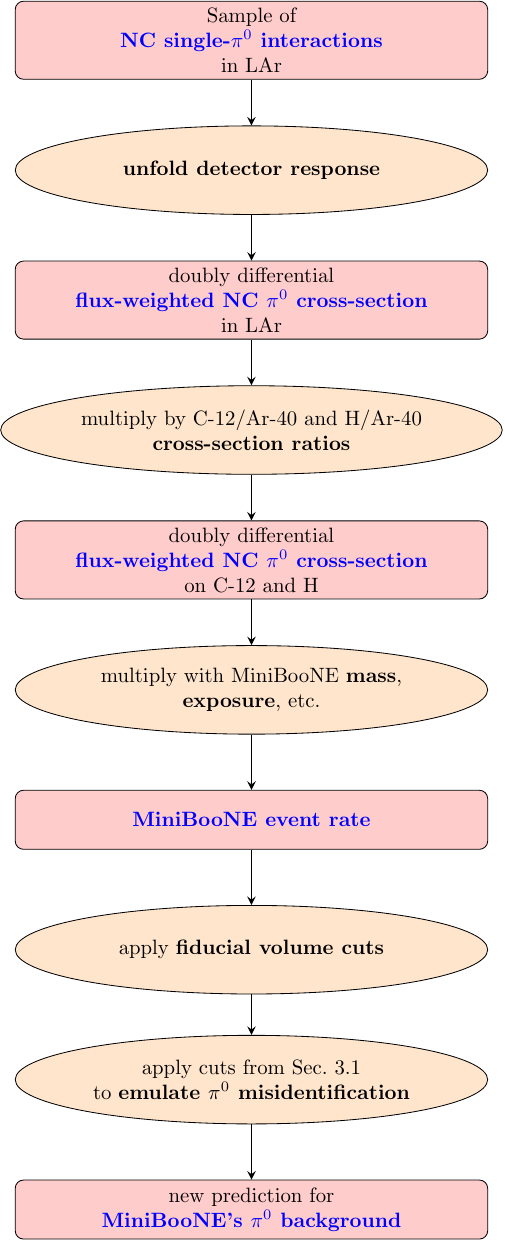}
    \caption{A possible strategy for predicting the $\pi^0$ background to MiniBooNE's $\nu_e$ appearance search based on the $\pi^0$ spectrum from a liquid argon detector and on the results from this paper.}
    \label{fig:flowchart-pi0-LAr}
\end{figure}

We now discuss the implications which the above musings on MiniBooNE's NC$\pi^0$ background have for liquid argon detectors. In particular, the parameterized description of the $\pi^0$/$e^-$ separation cut could be used to predict this background for MiniBooNE based on observations at MicroBooNE, SBND, and ICARUS, but independent of MiniBooNE's own data.  A possible strategy is outlined in \cref{fig:flowchart-pi0-LAr}: starting from well-reconstructed NC$\pi^0$ events in liquid argon, it should be possible to extract the flux-weighted cross-section as function of the pion's momentum and direction.  Multiplying by the theoretically predicted ratio of cross-sections on C-12 vs.\ Ar-40 and hydrogen vs.\ Ar-40, one can  then predict the rate of NC$\pi^0$ events in MiniBooNE. The cuts from \cref{subsec:CutApproach} can then be applied to describe the likelihood of a $\pi^0$ being misidentified as an electron. As shown in the preceding sections, these cuts are rather robust, and an excellent approximation to MiniBooNE's full likelihood analysis.

A background prediction based on this procedure would benefit from the large sample of well-reconstructed NC$\pi^0$ events that can be expected from the liquid argon detectors comprising Fermilab's short-baseline neutrino program. It is relatively -- though not completely -- robust against theoretical uncertainties, given that only a \emph{ratio} of cross-sections needs to be predicted, and that this prediction is mainly needed for neutrino energies large enough for nuclear effects to be subdominant.

\section{Summary}
\label{sec:Conclusions}

To summarize, we have presented some reflections on the $4.7\sigma$ excess of low-energy ($\text{few} \times \SI{100}{MeV}$) $\nu_e$-like events in MiniBooNE. In the first part of the paper, we have considered two-particle--two-hole (2p2h) interactions with final state radiation. In such events, only the final state photon would be visible to MiniBooNE and would mimic a $\nu_e$-induced electromagnetic shower. We have developed a simple toy model for 2p2h interactions with and without extra radiation, and while this model is certainly not suitable for precision calculations, it has proven useful in estimating the impact of such interactions on the MiniBooNE anomaly. As shown in \cref{fig:MBEvts_EQE}, we predict about 40 extra background events from this channel, which would reduce the significance of the anomaly by about $0.4\sigma$. While the contribution to the MiniBooNE excess from this class of events is not too dramatic, we hope that identifying this class leads to more detailed investigations, especially as we look forward to future measurements from the liquid argon SBN experiments.

We have then turned our attention to MiniBooNE's NC$\pi^0$ background. In doing so, we have constructed a phenomenological approach that quickly and faithfully reproduces MiniBooNE's ability to distinguish between electron-like (single-shower) and pion-like (two-shower) events. This approach is based off the kinematical quantities of the $\pi^0\to\gamma\gamma$ showers, subject to detector uncertainties on the energy/direction of the photons. We have demonstrated how, with an unrealistic level of data/Monte Carlo disagreement, significantly larger rates of events could pass these cuts and contribute to the MiniBooNE low-energy excess.

Our method for reproducing MiniBooNE's cuts can prove useful in scrutinizing other beyond-the-Standard-Model explanations of the MiniBooNE excess, including those that propose novel physics processes that lead to multiple-electron final states in the detector.

Additionally, we have compared how different neutrino event generators lead to different predictions for the NC$\pi^0$ background. We have observed that, if all generator predictions are normalized to MiniBooNE's measurement of $\pi^0$ production, then every generator other than \nuance predicts a significant upturn of NC$\pi^0$ events at low energies, exactly where the MiniBooNE excess occurs. While this upturn is not enough to account for the $4.7\sigma$ excess, it still highlights the importance of neutrino event generators in the search for new physics in neutrino facilities.

One aspect that we have not commented on so far is the impact that our results have on explanations of the MiniBooNE anomaly in terms of sterile neutrino oscillations \cite{Gariazzo:2017fdh, Dentler:2018sju, Moulai:2019gpi} or other physics beyond the Standard Model (see for instance Refs.~\cite{Dasgupta:2021ies,Acero:2022wqg} and references therein). In fact, any decrease in the number of excess events engendered by updated background predictions will move the favored regions in the sterile neutrino parameter space towards lower mixing angles. This should significantly reduce the tension between the MiniBooNE anomaly and null searches for muon neutrino disappearance. Simultaneously, the tension between MiniBooNE on the one side and the LSND and gallium anomalies on the other side may be modified, changing the likelihood that all anomalies have a common new physics explanation.

While our results indicate that the significance of the MiniBooNE anomaly may be slightly lower than previously thought, we note that the significance of the anomaly remains very high. In any case, it is clear that the ``Altarelli cocktail'' we propose here is still missing some ingredients -- either within the Standard Model or beyond. Fortunately, we have every reason to expect that the upcoming short-baseline experiments at Fermilab will reveal these secret ingredients.

\begin{acknowledgments}
It is a pleasure to thank Pedro Machado for valuable discussions and collaboration on early stages of this project, and Vedran Brdar for very useful feedback on an earlier copy of this manuscript. We have moreover benefited tremendously from exchanges with Omar Benhar and Ulrich Mosel. Finally, this work would not have been possible without innumerable in-depth discussions with members of the MiniBooNE collaboration, notably Janet Conrad, Bill Louis, Austin Schneider, and Mike Shaevitz, for which we are very grateful.
\end{acknowledgments}

\appendix
\section{Complete Expressions for cross-sections}\label{app:Expressions}

In this appendix, we provide complete expressions for the $2\to 3$ process of 2p2h$\gamma$ scattering, as well as for the squared 4-momentum transfer, $Q^2$, that enters the form factor for this process. The matrix elements for the diagrams in \cref{fig:FD}, $\mathcal{M}_1$ and $\mathcal{M}_2$, combine into the total squared matrix element via
\begin{align}
    \left|\mathcal{M}\right|^2 = \left|\mathcal{M}_1\right|^2
                               + \left|\mathcal{M}_2\right|^2
                               + \mathcal{M}_1\mathcal{M}_2^\dagger
                               + \mathcal{M}_2\mathcal{M}_1^\dagger \,.
\end{align}
The individual pieces in this expression may be expressed as
\begin{widetext}
\begin{align}
    \left|\mathcal{M}_1\right|^2
      &= -\frac{512c_w^4 e^2 G_F^2}{(p_2 k_\gamma)^2}
          \Big[ m_X^2 - (p_2 k_\gamma) \Big] \Big[
              (p_1(k_2 - k_\gamma + p_2)) \times (k_1(k_2 - k_\gamma + p_2)) \notag\\
      &\hspace{5cm}
            + (p_1 k_1)\left(k_2 k_\gamma - k_2 p_2 + p_2 k_\gamma - m_X^2\right)
          \Big] \,, \\[0.2cm]
    \left|\mathcal{M}_2\right|^2
      &= \frac{512c_w^4 e^2 G_F^2}{(k_2 k_\gamma)^2}
         \Big[ m_X^2 + (k_2 k_\gamma) \Big] \Big[
             - (p_1 (k_2 + k_\gamma + p_2)) \times (k_1(k_2 + k_\gamma + p_2)) \notag\\
      &\hspace{5cm}
             + (p_1 k_1) \left(k_2 k_\gamma + k_2 p_2 + p_2 k_\gamma + m_X^2 \right)
         \Big] \,, \\[0.2cm]
    2\mathrm{Re} (\mathcal{M}_1\mathcal{M}_2^\dagger)
      &= \frac{-512c_w64 e^2 G_F^2}{(p_2 k_\gamma)(k_2 k_\gamma)}
         \Big[ 2(p_2 k_2) + (p_2 k_\gamma) - (k_2 k_\gamma) \Big] \Big[
             (k_1 k_\gamma) (p_1 k_\gamma)                 \notag\\
      &\hspace{4cm}
           - (p_1 k_2 + p_1 p_2)(k_1 k_2 + p_2 k_1)
           + (p_1 k_1) (m_X^2 + p_2 k_2)
         \Big] \,,
\end{align}
\end{widetext}
where dot products between different four-momenta are implied. Furthermore, $G_F$ is the Fermi constant, $c_w$ is the cosine of the weak mixing angle, and $e$ is the electric charge. The total differential cross-section of 2p2h$\gamma$ scattering is then.
\begin{align}
    \frac{d\sigma}{dE_X^\text{cm} dE_\gamma^\text{cm} d\cos\theta_\gamma d\eta}
        = \frac{\left|F(Q^2)\right|^2}
               {16(2\pi)^4 (s - m_X^2)} \left|\mathcal{M}\right|^2 \,.
    \label{eq:DiffXSec}
\end{align}
It is given here in terms of the center-of-mass-frame energies of the two-nucleon system, $E_X^\text{cm}$, and the photon, $E_\gamma^\text{cm}$, the direction of the outgoing photon with respect to the incoming neutrino direction, $\cos\theta_\gamma$, and a second angle, $\eta$. The latter gives the orientation of the $(x,z)$ plane spanned by the incoming neutrino and the outgoing photon relative to the plane spanned by the outgoing particles (incoming particles are travelling in the $\pm z$ direction). See Ref.~\cite{FC5-manual} for a more detailed discussion of this parameterization of the geometry. In terms of the lab-frame neutrino energy, $E_\nu$, the squared center-of-mass energy appearing on the right-hand side of \cref{eq:DiffXSec} is given by $s = m_X^2 + 2E_\nu m_X$. As the differential cross-section is logarithmically divergent for $E_\gamma \to 0$, we impose a lower cutoff on $E_\gamma$ when integrating over phase space, as demonstrated in \cref{fig:XSec}.

The form-factor $\left|F(Q^2)\right|^2$ in the numerator of \cref{eq:DiffXSec} has been discussed in \cref{subsec:WithoutPhoton} (see in particular \cref{fig:FF}), where we extracted it by comparison of the 2p2h cross-section (without final-state radiation) in our toy model against the results of Ref.~\cite{Gallmeister:2016dnq}. In the two-body 2p2h process, $Q^2$ is a simple combination of four-momenta and can be expressed in terms of Mandelstam variables as $Q^2 = -t$. For the three-body 2p2h$\gamma$ process, however, the expression for $Q^2 = -(k_1 - p_1)^2$ is more complex. (Here, $k_1$ and $p_1$ are the outgoing and incoming neutrino 4-momenta, respectively.) In terms of the same kinematical variables as above, $Q^2$ is given by
\begin{widetext}
\begin{align}
    Q^2 &= \frac{s - m_X^2}{2\sqrt{s}E_\gamma^\text{cm}} \notag\\
        &\times \bigg( -\cos\eta \sin\theta_\gamma \sqrt{- \left(s + m_X^2 - 2\sqrt{s} E_X^\text{cm} \right)
                                             \left(s + m_X^2 - 2\sqrt{s}(E_X^\text{cm}+2E_\gamma^\text{cm})
                                                 + 4 E_\gamma (E_X^\text{cm}+E_\gamma^\text{cm})\right)}
                                               \notag \\
       &\quad + \cos\theta_\gamma \left(s + m_X^2 + 2(E_\gamma^\text{cm}-\sqrt{s})(E_X^\text{cm}+E_\gamma^\text{cm}) \right)
                                               \notag \\
       &\quad + 2 E_\gamma^\text{cm} \left(\sqrt{s} - E_\gamma^\text{cm} - E_X^\text{cm} \right) \bigg) \,.
    \label{eq:QsqFull}
\end{align}
\end{widetext}
Integrating this expression over $\cos\theta_\gamma$ and $\eta$ yields \cref{eq:QsqAvg}.

\section{Cut Values used in NC\texorpdfstring{ Single-Pion}{$\pi^0$} Analysis}
\label{app:CutTable}

For completeness, \cref{tab:CutTable} provides the cut values used for $\pi^0/e^-$ separation discussed in \cref{subsec:CutApproach} for the three different cut prescriptions we use. These are all derived based on \nuance MC events.
\begin{table}
\begin{center}
    {\footnotesize
    \caption{Threshold values for the $\pi^0$/$e^-$ separation cuts discussed in \cref{subsec:CutApproach}, derived using \nuance.\label{tab:CutTable}}
    \begin{filecontents*}{CutTable_v2.csv}
cola,colb,colc,cold,cole
0.140,0.160,0.334,0.520,1.000
0.160,0.180,0.283,0.441,1.046
0.180,0.200,0.244,0.381,1.091
0.200,0.220,0.237,0.366,1.098
0.220,0.240,0.203,0.315,1.140
0.240,0.260,0.181,0.281,1.168
0.260,0.280,0.165,0.256,1.188
0.280,0.300,0.150,0.232,1.209
0.300,0.320,0.140,0.217,1.221
0.320,0.340,0.136,0.209,1.227
0.340,0.360,0.112,0.174,1.260
0.360,0.380,0.118,0.183,1.250
0.380,0.400,0.113,0.174,1.257
0.400,0.420,0.110,0.170,1.261
0.420,0.440,0.106,0.163,1.267
0.440,0.460,0.100,0.155,1.274
0.460,0.480,0.105,0.162,1.268
0.480,0.500,0.096,0.148,1.281
0.500,0.520,0.091,0.140,1.288
0.520,0.540,0.089,0.138,1.289
0.540,0.560,0.084,0.130,1.297
0.560,0.580,0.083,0.128,1.299
0.580,0.600,0.095,0.147,1.281
0.600,0.620,0.087,0.134,1.293
0.620,0.640,0.089,0.138,1.290
0.640,0.660,0.092,0.142,1.286
0.660,0.680,0.083,0.128,1.298
0.680,0.700,0.091,0.140,1.287
0.700,0.720,0.085,0.130,1.296
0.720,0.740,0.074,0.115,1.310
0.740,0.760,0.095,0.147,1.282
0.760,0.780,0.093,0.144,1.285
0.780,0.800,0.089,0.138,1.290
0.800,0.820,0.091,0.142,1.287
0.820,0.840,0.090,0.141,1.288
0.840,0.860,0.068,0.104,1.320
0.860,0.880,0.090,0.141,1.289
0.880,0.900,0.088,0.139,1.291
0.900,0.920,0.081,0.127,1.301
0.920,0.940,0.076,0.120,1.307
0.940,0.960,0.090,0.143,1.289
0.960,0.980,0.072,0.112,1.313
0.980,1.000,0.080,0.126,1.303
1.000,1.020,0.095,0.151,1.282
1.020,1.040,0.083,0.131,1.299
1.040,1.060,0.095,0.153,1.283
1.060,1.080,0.056,0.088,1.336
1.080,1.100,0.109,0.181,1.264
1.100,1.120,0.101,0.168,1.275
1.120,1.140,0.156,0.279,1.207
1.140,1.160,0.081,0.130,1.301
1.160,1.180,0.095,0.155,1.283
1.180,1.200,0.091,0.151,1.288
1.200,1.220,0.102,0.174,1.273
1.220,1.240,0.100,0.167,1.277
1.240,1.260,0.097,0.163,1.280
1.260,1.280,0.115,0.202,1.257
\end{filecontents*}
\pgfplotstabletypeset[
    col sep=comma,
    string type,
    columns/cola/.style={column name=$E_{\rm vis.,min.}$ [GeV], column type={|c}},
    columns/colb/.style={column name=$E_{\rm vis.,max.}$ [GeV], column type={|c}},
    columns/colc/.style={column name=$r_{\textsc{Diagonal}}$, column type={|c}},
    columns/cold/.style={column name=$r_{\textsc{Circle1}}$, column type={|c}},
    columns/cole/.style={column name=$r_{\textsc{Circle0}}$, column type={|c|}},
    every head row/.style={before row={\hline
            \multicolumn{2}{|c||}{Visible Energy Range} & \multicolumn{3}{c||}{Cut Values}\\\hline
        },after row=\hline\hline},
    every last row/.style={after row=\hline},
    ]{CutTable_v2.csv}
    }
\end{center}
\end{table}

\section{Results using Out{-}of{-}the{-}box Generators}
\label{app:OutOfBox}

In our discussion of MiniBooNE's NC$\pi^0$ background in \cref{sec:NCPi0}, specifically in the calculations that produced \cref{fig:GeneratorComp}, we started by re-normalizing the expected NC$\pi^0$ event rate from an MC generator to match the total number of (correctly identified) $\pi^0$ spectra presented in Ref.~\cite{MiniBooNE:2009dxl}. The corresponding reweighting facors were shown in \cref{tab:Reweight}. 

\begin{figure}
    \centering
    \includegraphics[width=\linewidth]{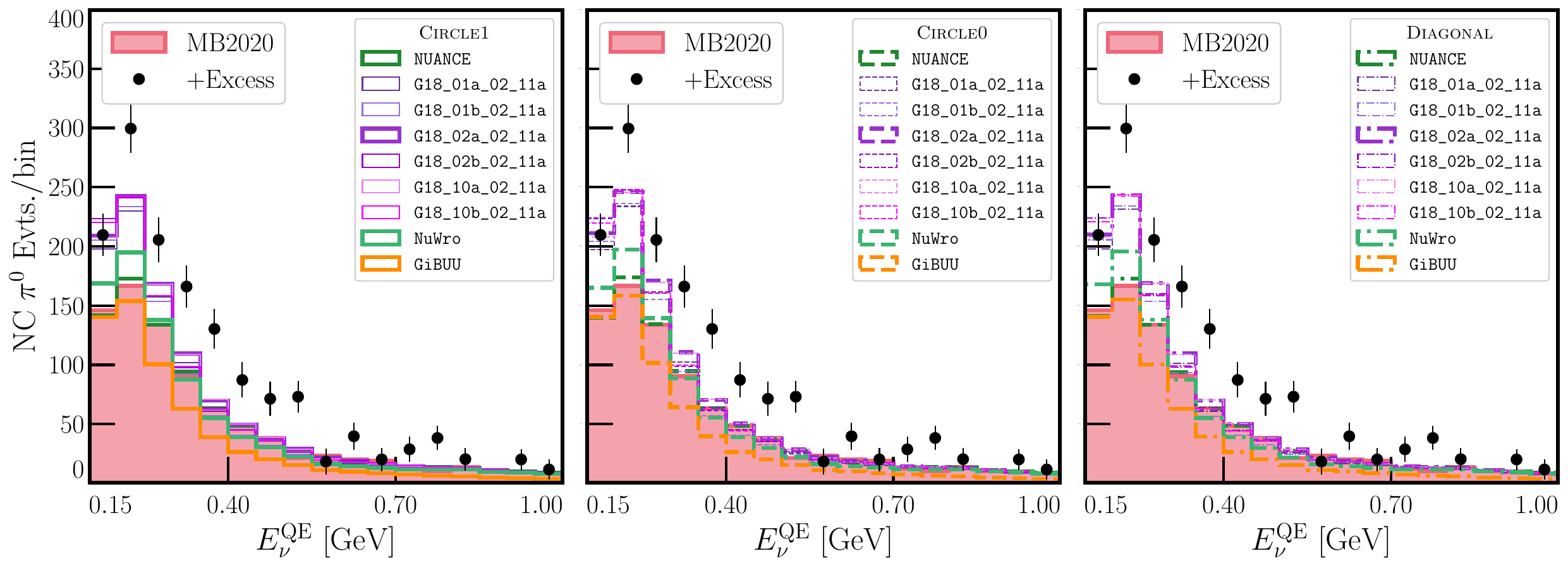}
    \caption{Distributions of NC$\pi^0$ background events in MiniBooNE (neutrino mode) predicted by different Monte Carlo generators. This figure is similar to \cref{fig:GeneratorComp} in the main text, but here, we do \textit{not} normalize the events from the MC generators based on the MiniBooNE measurement of NC$\pi^0$ events in Ref.~\cite{MiniBooNE:2009dxl}.}
    \label{fig:GeneratorCompOOB}
\end{figure}

We repeat these analyses here, but now using the generators ``out of the box,'' i.e.\ \textit{not} re-normalizing their predictions. We note that this means re-deriving cut thresholds for electron/pion separation based on an out-of-the-box \nuance prediction that is ${\sim}75\%$ of the one we studied in the main text. We then apply these cuts to the various neutrino event generator samples without any normalization according to the measured NC$\pi^0$ rate from Ref.~\cite{MiniBooNE:2009dxl}. The result is presented in \cref{fig:GeneratorCompOOB}. Because the generators' $\pi^0$-production rates have not been normalized to the MiniBooNE measurement, we now observe much more spread between their predictions, ranging from as low as 493 NC$\pi^0$ events misidentified as CCQE $\nu_e$ (from the \texttt{GiBUU} generator using the \textsc{Circle1} cut) to as high as 874 (for \texttt{GENIE} tune \texttt{G18\_02a\_02\_11a} using the \textsc{Circle0} approach). In general, \texttt{GENIE} predictions are characteristically high. 

Thus, we conclude by saying that, if the $\texttt{GENIE}$ Monte Carlo predictions were to be trusted blindly, the MiniBooNE low-energy excess would be far less significant than previously stated.  In reality, however, the data-driven methods used in the main text (and also by the MiniBooNE collaboration) are more robust and should be trusted more.

\bibliographystyle{apsrev4-1}
\bibliography{references}

\end{document}